\documentstyle[preprint,eqsecnum,aps,epsfig]{revtex}

\tightenlines

\newcommand{\cs}{\'{c}}

\newcommand{\beq}{\begin{equation}}
\newcommand{\eeq}{\end{equation}}
\newcommand{\bdm}{\begin{displaymath}}
\newcommand{\edm}{\end{displaymath}}
\newcommand{\beqa}{\begin{eqnarray}}
\newcommand{\eeqa}{\end{eqnarray}}
\newcommand{\beqab}{\begin{eqnarray*}}
\newcommand{\eeqab}{\end{eqnarray*}}
\newcommand{\de}{\partial}

\newcommand{\Ms}{M_*}

\def\nn{\nonumber}

 \def\@makefnmark{\hbox to 0pt{$^{\@thefnmark}$\hss}}  

\def\R4{R}
\def\M{M_*}
\def\msm{M_{\rm SM}}
\def\mpl{M_{\rm P}}
\def\pr{\partial}
\def\pl{{\overline M}}

\def\d{\delta(y)}

\newcounter{saveeqn}%

\begin{document}


\preprint{NYU-TH-01/06/03;~~TPI-MINN-01/26}

\title{Scales of Gravity}

\author{Gia Dvali$^1$ \footnote{e-mail:gd23@nyu.edu}, Gregory 
Gabadadze$^2$
 \footnote{e-mail: gabadadz@physics.umn.edu}
Marko Kolanovi\cs$^1$ \footnote{e-mail: mk679@nyu.edu}, Francesco 
Nitti$^1$ 
\footnote{e-mail: fn230@nyu.edu} }
\address{$^1$Department of Physics, New York University,
New York, NY 10003\\
$^2$Theoretical Physics Institute, University of Minnesota, 
Minneapolis, MN, 
55455}
\date{\today}
\maketitle
\begin{abstract}

We propose a framework in which the quantum gravity scale 
can be as low as $10^{-3}$ eV. The key assumption
is that the Standard Model ultraviolet cutoff is much
higher than the quantum gravity scale.
This ensures that we observe conventional  weak gravity.  
We construct an
explicit brane-world model in which the brane-localized
Standard Model is coupled to strong 5D gravity of infinite-volume
flat extra space. Due to the high ultraviolet scale, the 
Standard Model fields generate a large graviton 
kinetic term on the brane. 
This kinetic term ``shields'' the Standard Model from
the strong bulk gravity. As a result, an observer on the brane sees
weak 4D gravity up to astronomically large distances
beyond which gravity becomes five-dimensional.
Modeling quantum gravity above its scale
by the closed string spectrum we show that the 
shielding phenomenon
protects the Standard Model from an apparent phenomenological 
catastrophe  due to the exponentially large number of light 
string states. The collider experiments, astrophysics, cosmology  and
gravity measurements {\it independently}
point to the same lower bound on the quantum gravity scale,
$10^{-3}$ eV. 
For this value the model has experimental signatures both for colliders
and  for sub-millimeter gravity measurements.
Black holes reveal certain interesting  properties in this framework.

\end{abstract}

\newpage

\section{Introduction}
\setcounter{equation}{0}

One of the great mysteries about gravity is its 
inexplicable weakness compared to all the other known forces of Nature.
For instance, the magnitude of the 
Newton  gravitational force between  two protons 
is $10^{36}$ times smaller  than the magnitude 
of  the Coulomb force
between them.  In the language of the low energy
field theory the Newton  force is mediated via the exchange of a 
virtual massless spin-2 particle, the graviton $h_{\mu\nu}$,
which couples to the matter energy-momentum tensor $T_{\mu\nu}$ as 
follows:
\begin{equation}
{1 \over \mpl}~h_{\mu\nu}~T^{\mu\nu}~.
\end{equation}
The corresponding coupling constant has the dimensionality and magnitude 
of the Planck mass, $M_{\rm P}\sim 10^{18}$ GeV.
Therefore, the dimensionless
ratio that governs the strength of the Newton  force 
between the two protons is
\begin{equation}
 \alpha_{\rm gravity}~ \sim ~ {m_{\rm proton}^2 \over M_{\rm P}^2}~.
\end{equation}
This has to be confronted with the electromagnetic coupling constant
$\alpha_{EM} = 1/137$  when one compares the Newton  and Coulomb forces.
In this language gravity is weak  because of the huge  
value of  $M_{\rm P}$ compared to the particle masses.
 
What is the meaning of the scale $M_{\rm P}$?
For some time the answer to this 
question has been known as what is called  the ``Standard Paradigm''.
According to this paradigm $M_{\rm P}$ is the fundamental scale 
of quantum gravity, i.e., the scale at which 
the classical theory of gravitation, as we know it,
should cease to be valid description of Nature.
This is based on the  assumption that the effective 
low-energy description in terms of Newtonian or Einstein 
gravity breaks down at the Planck energies 
$\sim M_{\rm P}$ or, equivalently, at  the 
Planck distances $l_{\rm P} = 1/M_{\rm P} \sim 
10^{-33}$ cm, and therefore, quantum gravity effects become 
important. On the other hand, the validity of 
the Newtonian interactions is experimentally measured 
only for distances bigger than  
$\sim 0.2$ millimeter \cite {measurements}.
The standard paradigm uses the following assumptions:

\begin{itemize}

{\item  It assumes that nothing is happening with gravity all the way 
down to the distances of order $10^{-33}$ cm, or equivalently,
all the way up until the energies of order $M_{\rm P}$. Therefore,
it extrapolates the known experimental result of the gravitational
measurements by 31 orders of magnitude.}

{\item It also assumes that $M_{\rm P}$ is a natural field 
theory cut-off not only for gravity, but for the whole particle 
theory including the Standard Model (SM) and its 
extensions (e.g., Grand Unified Theories (GUT), etc).}

\end{itemize}

The existence of an energy ``desert'' stretched over 17 orders of 
magnitude gave rise to the  hierarchy problem.
As seen from such an angle, the assumptions of the standard paradigm look 
somewhat unnatural.  After all, if we think of known field theory
examples  such as, e.g.,  
electrodynamics, quantum effects become important at scales at which 
the coupling is still very weak and perturbation theory is valid. 
There is no 
{\it a priory} reason for a theory to ``wait'' until the classical
interactions blow up, in order for the quantum effects to become 
important.  From this perspective, 
it is natural to question this paradigm and ask 
whether the gravitational cut-off might  be much lower in reality.

This was done in Refs. \cite{add},\cite {Lykken0}. 
The approach of \cite{add} was mainly motivated by the hierarchy problem.
It assumed that the quantum gravity scale, referred hereafter as $M_*$,
and  the field theory cut-off of SM and its possible extensions,
referred below  as $\msm$, 
were around the electroweak scale, i.e.,  $\sim$ 1~TeV.

Although the quantum gravity scale $M_*$ is very low 
in the framework of \cite{add}, 
it nevertheless shares one common 
assumption with the standard paradigm: in both these approaches 
$M_* \sim \msm$.

In the present paper we relax this assumption and show that the quantum 
gravity scale $M_*$ can be much lower than the 
field theory cut-off $\msm$  without conflicting with any of the 
existing laboratory, astrophysical or cosmological bounds.

We consider a theory in which gravity, becoming ``soft'' 
above the scale $M_*$, 
is coupled to the SM which remains consistent field theory
up to the scale $\msm \gg M_*$. 
We show that despite the small 
quantum gravity scale, the weakness of an observable 
gravity is guaranteed by the high cut-off of the SM. 
In this framework, the Planck mass $\mpl$  
is not a fundamental scale but is rather a derived
parameter.  The large value of the observable $M_{\rm P}$ is 
determined by $\msm$ rather than $M_*$.  

In the present work we will mostly be concentrated
on a brane-world model of Ref. \cite {dgp}. However,
the similar consideration  is applicable in a conventional
four-dimensional theory which we will discuss in section II.  
As an example of this phenomenon 
in four dimensions  consider the following action
(we will clarify the origin of this action in section II)
\beqa
S~=~\int d^4x~\sqrt{|g|}~\left (\mpl^2~R~+~\sum_{n=1}^{\infty} 
c_n~{R^{n+1}\over \M^{2(n-1)}}  \right )~.
\label{ex}
\eeqa
Here $\M\ll\mpl$, $R$ denote the four-dimensional Ricci scalar 
and higher powers of $R$  denote all possible higher-dimensional 
invariants ($c_n$'s are some constants). 
Suppose that this gravitational theory is coupled
to the SM fields in a conventional way. Let us consider 
the gravitational interactions of the SM 
particles at the distances much bigger than $1/\M$. 
Since the coefficient in front of the Einstein-Hilbert term
is $\mpl^2$, the gravitational coupling  
of the SM fields is proportional to $1/\mpl$ and is very weak.
However, the higher derivative terms in (\ref {ex}) 
become important and the gravitational 
self-interactions become of order 1 at the distances 
below $1/\M$. Thus, the effective low-energy approximation 
to gravity  breaks down at an  energy scale of order $\M$. 
Above this scale quantum gravity corrections 
should be taken into account. If these corrections 
are ``soft'', i.e., if they do not lead to 
an effective increase of the coupling of the SM to  gravity,
then this breakdown of the effective gravitational theory
would not be observable in any present day high-energy 
particle physics experiments which are  
insensitive to the effects of the gravitational 
strength (defined by $1/\mpl$).
Therefore, the gravitational interactions for the SM 
particles  with energies above $M_*$ will still remain 
very weak compared to the  SM gauge interactions.

The idea that effective field theory description
of gravitation can brake down at distances smaller than,
but close to, a millimeter was put forward by R. Sundrum 
\cite {sundrum}. Motivated by toy QCD examples 
he discussed the ``compositness scale'' (string scale) 
for gravity at an inverse millimeter with the purpose 
to solve the cosmological constant problem. 

In theories with a super-low quantum gravity scale
(i.e., when $\M\ll \msm$) there are at least three 
important issues to be addressed. 
First, it is not clear {\it a priory}
why the gravitational constant is determined by the scale
$1/\mpl^2$ if its field-theory description breaks down at
the scale $\M$; one would rather expect in this case 
that the gravitational coupling is determined by $1/\M^2$. 
As a resolution of this puzzle, we will show that the 
loops of the SM with the UV scale $\msm$ renormalize 
the strong gravitational coupling 
$1/\M^2$ to make it weak, that is $1/\mpl^2$. 
The second issue deals with the
huge hierarchy between the scales $\M$ and $\msm$.
Although this hierarchy is stable by itself
(i.e., the hierarchy is technically natural), 
still it is desirable to have some  dynamical realization 
for it. We will argue in section III that 
brane-world scenarios can offer such a realization. 
In fact, we present a toy brane-world model which has a 
string theory realization and naturally gives rise to 
the hierarchy $\M\ll \msm$.  
The third issue concerns the assumed ``softness''
of gravity above the scale $\M$.  
In order to understand this issue in more details
one should have a model for quantum gravity. 
At present, a candidate for quantum gravity is string theory
which is formulated in higher dimensions.
Therefore, the latter two issues motivate us to 
go to higher dimensional theories. An attractive possibility 
for this, as we mentioned above, is the brane-world scenario. 

A 5D brane-world framework which  explicitly realizes 
this idea was introduced in Ref. \cite{dgp}. 
The model has a 3-brane embedded in 
flat uncompactified  5-dimensional space where gravity propagates. 
The SM fields are assumed to be confined to the brane. 
The field theory cut-off  of the bulk gravity 
is $\M$.  The  effective world-volume theory on the brane is 
a field theory with a very high cut-off $\msm$. 
In this framework $\msm \gg \M$ (see 
discussion of an example of this type in section III).
The world-volume theory is coupled to  the bulk gravity. 

Despite the existence of an infinite-volume 5D space with a
strong gravitational constant proportional to $1/\M^3$, 
an observer on the brane 
measures  the 4D weak gravity with the conventional  Newton coupling 
$G_N = 1/16\pi M_{\rm P}^2$ within the following intermediate distances: 
\beq
M_*^{-1}~ \ll ~r~ \ll ~r_c~\equiv ~{M_{\rm P}^2\over M_*^3}~.
\label{distances}
\eeq
However, for the distances $r \gg r_c $ gravity becomes five dimensional.
The reason for such an unusual behavior is as follows. 
Consider the renormalization of the graviton kinetic 
term due to matter loops on the brane (Fig.1). 
The diagram  with massive states in the loop 
gives rise to the renormalization of the 
4D graviton kinetic term 
which is dominated by the states with the masses close to the 
SM cut-off $\msm$.  The resulting term in the action has the form
\beq
S_{\rm ind}~=~\gamma\,\msm^2\int d^{4}x~
\sqrt{|{g}|}~{\R4}(x)~,
\label{term}
\eeq
where $g$ is the induced metric on  the brane  and ${\R4}(x)$ is the 
corresponding four dimensional Ricci scalar.
The coefficient of this term, $\gamma$, depends on
the number of states and particle content of 
the SM running in the loop.
The multiplier in (\ref {term}), that is  $\gamma \msm^2$,
 should be set equal to 
the 4D Planck mass  $\mpl$  due to phenomenological requirements. Indeed,
we will see that  the effective Newton constant measured by the 
brane observer is equal to the inverse value of this constant, 
$G_N=1/(16\pi\gamma \msm^2)$.
Therefore, the Planck scale in this framework is 
not a fundamental quantity but is rather a derived scale 
which is related to the 
SM cutoff $\msm$ (or to the GUT cutoff) 
and its particle content.
   
The crucial point is that the 4D Newton constant is set by 
the cut-off of the SM which is much bigger than $M_*$.
The induced  term plays a  crucial role in what follows.
It ensures that a  brane observer measures  the weak 4D gravity at 
distances $r\ll r_c$. For the values of $M_*$ that are of 
our interests, $r_c$ is astronomically large.
Thus the crucial question  follows: what is the lower bound on $M_*$? 

It was already noticed  in Ref. \cite{dgkn} that in the effective 
field theory picture there is  no phenomenological constraint 
that would forbid $\M$ to take any small value all the way down to 
$10^{-3}$ eV. 
The reason is as follows \cite {dgkn}:  
Due to the induced term (\ref{term}) gravity on a brane becomes more and
more four-dimensional as we increase energy 
and all the high energy reactions with graviton 
emission from the SM states proceed as in the conventional 
weak 4D gravity. Thus the rigid SM ``shields'' itself 
against the strong bulk gravity. 
Nevertheless, as it was suggested in Ref. \cite{adgp},
the low value of $M_*$ may still manifest itself in 
gravitational measurements at scales $r\ll M_*^{-1}$.
This constrains  the value of $M_*^{-1}$ to be smaller than the  
distance at which the gravitational interactions 
between static sources are presently measured, i.e., 
$M_*^{-1} \lesssim 0.2$ mm \cite{measurements}. 
Therefore,  $\M > 10^{-3}$ eV.

In the discussions above the graviton momenta were effectively 
cut-off at $M_*$ in all the high energy SM processes due to the lack 
of the knowledge of the precise theory of 
quantum gravity above the scale $M_*$. 
However, the behavior of gravity above this scale may dramatically 
change the conclusions. At present, the  construction of a
realistic brane world-model from string theory 
which would possess all the desired phenomenological 
properties is difficult. 

For this reason, we attack this problem 
from a somewhat pragmatic point of view. 
We assume  that above the scale $M_*$ gravity is described by  
a model which mimics in many respects  
the crucial properties of string theory. 
In particular, we assume that the bulk theory has 
the mass spectrum and multiplicity 
which is similar to that of a closed bosonic  
string theory in critical dimension (neglecting tachyon).
Doing so, we will be able to construct  a toy 
model with some of the crucial features of string theory  
(most importantly, the huge multiplicity of states) 
which would naively invalidate any proposal 
with a low quantum gravity scale.   

In the present model the massive stringy states become important 
at the scale $\M$. Furthermore, we assume that the  
stringy tower of bulk states couples to the fields of the 
Standard Model which are localized on a brane. 
Such a set-up, although  far 
from being a self-consistent string theory, nevertheless serves our 
purpose  of putting phenomenological constraints on $M_*$ and studying 
possible signatures.

We discover that the lower bound on $M_*$ is still  $10^{-3}$ eV.
This, surprisingly enough, comes from several  
rather independent considerations: 

\begin{itemize}

{\item The model, as we discussed, predicts the modification of  
Newtonian gravity at distances $r< M_*^{-1}$. 
This constrains $1/\M$ to be smaller  than $\sim 0.2$ mm, i.e.,
$\M$ to be bigger than $10^{-3}$ eV.}

{\item Collider phenomenology puts the same constraint on 
$\M$  since the rate of the production
of the bulk stringy Regge recurrences would become 
significant at the energies of order 
$\sqrt {\M \mpl}$ which would be less than a TeV  if $\M$ 
was smaller than $10^{-3}$ eV.} 

{\item Astrophysical bounds arising  from constraints 
on the rate of star cooling due to the emission of stringy Regge 
states put the same bound on $\M$.} 

{\item The Hagedorn type phenomenon 
can strongly affect  the early universe 
and in particular the big bang nucleosynthesis.
These cosmological considerations also constrain 
$\M$ to be bigger than $10^{-3}$ eV or so.}

\end{itemize}

For the value  of $M_*$ which saturates the bound $10^{-3}$ eV, 
the model has a number of very distinctive experimental 
signatures including the deviation from the Newtonian gravity at 
sub-millimeter scales, as well as the collider signatures due to the 
production of stringy Regge recurrences  
with the mass gap of the size of an inverse millimeter. 
Surprisingly enough, these prediction are somewhat similar to 
those obtained in the  models of Ref. \cite{add} with two
sub-millimeter extra dimensions. However, both the 
modification of Newtonian potential as well as the spectrum 
of missing energy in collider experiments are different.

In the present scenario the behavior of  
black holes is rather peculiar. Elementary particles heavier 
than $M_*$ can turn into  long-lived black holes if emitted 
from the brane into the bulk.

One interesting range for the parameter $\M$ is  
$\M ~\sim ~10$ MeV.  In this case, 
our model predicts in addition the 
modification of gravitational laws at  scales  
comparable with  the present cosmological 
horizon. This gives rise to a possibility 
to accommodate an accelerated 
4D Universe \cite{cedric} which is in agreement 
with the recent Supernovae and Cosmic Microwave Background 
observations \cite {cedric1}. The remarkable feature of this scenario is that
is does not require a small nonzero cosmological constant, 
instead, the acceleration takes place due to the 
presence of an infinite volume fifth dimension.

One more attractive feature of the present scenario is that
the SM fields are confined to the brane, and, therefore,
if supplemented by low-energy supersymmetry, 
the conventional logarithmic gauge coupling unification of a 4D 
theory \cite {UNIFICATION} holds unchanged. 

Furthermore, since quantum gravity in the present model 
becomes important at a millimeter, 
it is natural to explore the idea of Ref. \cite {sundrum}
on the cosmological constant in this context \cite {adgp}. 
This will be discussed in Ref. \cite {adgp}.   

The paper is organized as follows:  
In section II we discuss a 4D theory with 
no branes or extra dimensions.
We show how the scale of quantum gravity can 
be much smaller than the SM ultraviolet cutoff.
Moreover, we show how the SM fields ``shield''
themselves from strong gravity. 
In section III we describe the basic ingredients of the 
5D brane-world model. 
In addition, we propose a mechanism for localization 
of massive fields in the model; we also 
study the tensorial structure of the graviton propagator. 
In section IV we describe qualitatively why
the presence of light 
Regge recurrences in a theory with low $\M$ 
cannot affect strongly the 4D physics on the brane.
In section V we develop  a model which mimics 
basic properties of the string spectrum. We show how
the SM can ``shield'' itself from the huge multiplicity of 
the Regge states. 
In section VI we study  high-energy, astrophysical
and cosmological constraints on $\M$ which arise due to the 
high multiplicity stringy Regge states.
Section VII discusses 
some curious aspects of black hole physics in the present context. 
In section VIII we study the processes of baryon number violation
due to quantum gravity effects. Conclusions are given in section IX.
Some useful derivations and formulas are collected in Appendix.

\section{ A Four-Dimensional Example}

Before we discuss the five-dimensional brane model, 
we are going to show in this section how the gravitational 
coupling constant can be determined  by
``non-gravitational''  physics already  
in a simple  four-dimensional  example. 

Let us consider  a model with the following two scales:
\beqa
\M~\ll~\msm~.
\label{scales}
\eeqa
Here we assume that the SM cutoff $\msm$ can be as large as
the conventional GUT scale.
In the latter case one  
needs to stabilize the Higgs mass against radiative corrections.
Therefore, above 1 TeV the Standard Model should be embedded in some 
bigger theory (supersymmetry, extended technicolor or 
something else). In the paper we use 
for convenience the name Standard Model for this theory. 

Note that the hierarchy between the
scales $\M$ and $\msm$ (\ref {scales}) 
is {\it stable}; this is  similar to the stability of 
the QCD scale with respect to the electroweak scale.

Consider the following gravitational action
\beqa
S_G~=~\int d^4x~\sqrt{|g|}~\left (\M^2~R~+~ 
~\sum_{n=1}^{\infty} 
c_n~{R^{n+1}\over \M^{2(n-1)}}  \right )~,
\label{Sg}
\eeqa
where $R^n,~~n=2,3,...$, stand for all possible
higher derivative curvature invariants.
So far the  only scale in this model is $\M$. 
Since this scale is small, i.e., $\M\ll 10^{19}~{\rm GeV}$, 
the self-interactions of gravitons are strong.
The corresponding Newton constant is 
defined as follows: 
\beqa
G_*~=~{1\over 16~\pi~\M^2}\,.
\label{Gstar}
\eeqa
Furthermore, the effective field-theory description of gravity 
in (\ref {Sg}) ceases to be valid for energies above $\M$.
 
As a next step, let us couple the Standard Model fields to 
the gravity described by the action (\ref {Sg}). 
For this we introduce the action of the SM fields:
\beqa
S_{\rm SM}~=~\int~d^4x~\sqrt{|g|}~{\cal L}_{\rm SM}(\Psi,~\msm)~,
\label{Ssm}
\eeqa
where $\msm$ is the ultraviolet  cutoff of 
particle physics described by (\ref {Ssm}) and $\Psi$ 
collectively denotes all the SM fields. 
The total action we deal with is the sum:
\beqa
S~=~S_G~+~S_{\rm SM}~.
\label{SSS}
\eeqa
Gravity in (\ref {Sg},\ref {SSS}) is considered 
as an effective low-energy 
field theory up to  energies  of order $\M$. 
The SM, on the other hand, is supposed to be treated as a quantum field 
theory up to  the scale  $\msm$. This is the classical picture. 

The crucial point is that at the quantum level 
the Standard Model loops renormalize the gravitational 
action (\ref {Sg}). This renormalization is due to perturbative 
\cite {Capper} as well as nonperturbative \cite{Adler}
SM contributions.  
For the illustrative purposes consider a one-loop 
polarization diagram with two external graviton
legs and only SM heavy particles in the loop
\footnote{For simplicity of arguments 
we do not discuss here the other 
diagrams in the same order in which two graviton lines 
join at the same point.}. 
Let us set the momenta in the 
graviton external legs to be smaller than $\M$ so that 
(\ref {Sg}) provides valid classical description of
external graviton lines. On the other hand, the 
momentum in the loop in which the SM fields are running 
can take any value from zero all the way up to $\msm$.
As a result, this diagram gives rise to the 
renormalization of the graviton kinetic term \cite {Capper} 
which generically is determined by  the mass square 
of the heaviest SM particle in the loop 
(the latter we set to be of the order of $\msm^2$).
On the other hand, the similar diagrams with the 
gravitational lines in the loop cannot be calculated within the 
effective field theory approximation given by (\ref {Sg}). However,
given the assumption of the ``softness'' of gravity above
$\M$ these corrections become sub-dominant 
(see detailed discussions below and in section IV).   
In general, while dealing with the renormalization of the 
graviton kinetic term, one should take into account nonperturbative
SM contributions as well. All these contributions can be 
summarized by adding to the total action (\ref {SSS}) 
the following induced terms:
\beqa
\Delta S_{\rm ind} ~=~ M_{\rm ind}^2~\int~d^4x~ 
\sqrt{|g|}\left (~R~+~
{\cal O}\left ({R^2 \over \msm^2}\right )~+...\right ),
\label{loop2}
\eeqa
where the induced scale is defined as follows \cite{Adler,Zee}:
\beqa
M_{\rm ind}^2~=~{i\over 96}~
\int~d^4x~x^2~\left [\langle 0|T~T^\mu_\mu(x)~
 T^\nu_\nu(0) ~|0\rangle~-~
(\langle 0|T^\mu_\mu(0)~|0\rangle)^2 \right ]~,
\label{ind2}
\eeqa
and $T^\mu_\mu$ denotes the trace of the energy-momentum 
tensor of the fields of the particle theory (Standard Model)
\footnote{In the one-loop approximation 
scalars and spin-1/2 particles give rise to a positive
contributions to the induced kinetic term 
while gauge bosons lead to negative 
terms. We will assume that in the present model 
the overall sign of $M_{\rm ind}^2$ is positive.
Moreover, through the paper 
we neglect (i.e., we fine-tune to zero) the 4D cosmological constant
which is also induced by loops.}. 
Generically $M_{\rm ind}^2$ is expected to be of the order 
of the cut-off of SM, that is $M_{\rm ind}^2~\sim~\msm^2$.

The higher derivative terms which are also induced via the 
loop diagrams are suppressed by powers of $\msm$.
Since the latter scale 
is much bigger than $\M$,  we can neglect these higher 
derivative terms in comparison  with the ones which are 
suppressed by the smaller scale $\M$ and are   
already present in (\ref {Sg}).  

Therefore, the total action takes the form:
\beqa
S_{\rm total} ~=~ S_G~+~S_{\rm SM}~+~\Delta S_{\rm ind}~.
\label{tot3}
\eeqa
The net result is that due to the induced term 
the coupling of gravity to the SM fields is 
renormalized. In fact, this coupling becomes weaker.
The physical interpretation of this phenomenon will be
given at the end of this  section. 
Here we write the  resulting Newton constant:
\beqa
16~\pi~G_N~=~{1\over \M^2 ~+~M_{\rm ind}^2} ~\simeq 
~{1\over M_{\rm ind}^2}~\propto~{1\over \msm^2}~.
\label{loopaction}
\eeqa
For phenomenological reasons we have to put 
$M_{\rm ind}^2 \simeq \mpl^2$. Therefore, 
the Planck scale is a derived parameter and 
is completely defined by the content and dynamics of 
the corresponding particle physics theory.

Let us now turn to the higher derivative terms.
As before, they are suppressed by the scale $\M$:
\beqa
\sqrt{|g|}~ {R^{n+1} \over ~~\M^{2(n-1)}}~.
\label{hd}
\eeqa
As a result, the effective field theory approximation to gravity 
breaks down at the scale $\M$. 

Based on these considerations we can draw the 
following conclusions:
\begin{itemize}

\item{The matter fields are coupled to gravitons very weakly,
via the ordinary 4D Newton  constant $G_N=1/16\pi\mpl^2$.} 

\item{At low momenta, i.e., for $p\ll \M$, gravity couples to itself 
via the ordinary Newton  constant $G_N$. However, at 
high momenta, $p\geq \M$, the higher derivative terms in (\ref {Sg}) 
become important and there are additional contributions of the type
$p^n/\M^n$ in the graviton self-interaction vertices.}

\item{For low momenta, $p\ll \M$, the graviton propagator is that 
of a normal 4D massless particle. However, if $p\geq \M$ the 
propagator is modified by higher derivative terms.}

\end{itemize}

As we discussed before, we assume that 
gravity becomes ``soft'' above the scale $\M$ so that  
the coupling of matter fields to gravity remains weak
at any reasonable energies below $\msm$. 
From the practical point of view this means that 
the ``softening'' of gravity due to quantum effects
could be modeled  by some kind of formfactors in the 
gravitational vertices and propagators
(see sections IV, V). 

What we have seen in this section is that the SM
particles renormalize their own  gravitational couplings  
and make it weaker. 
The mass squared parameter in front of the Ricci scalar in the 
Einstein-Hilbert action is similar in this respect to 
the Higgs mass in the Standard Model: no matter how small
it is in the classical theory, the quantum loops drive its value
all the way up to the corresponding 
ultraviolet cutoff.

Let us try to understand this phenomenon
in terms of a simple physical picture. Suppose 
there is a single heavy scalar field which has a mass of the 
order of $\msm$ and which is the only state 
that  runs in the SM loop. The corresponding (additive)  
renormalization of  the gravitational constant is 
\beqa
\msm^2~{\rm log}\left ({\msm^2\over \mu^2}\right)~.
\eeqa
Here $\mu$ is the energy scale (normalization point) 
in the SM process which is 
less than  $\msm$ and bigger  than the  infrared cutoff of the SM.
Thus, the renormalization of the 
gravitational coupling is proportional to the  
mass square of the particle. 
This can be understood as follows.
Consider a heavy static source the gravitational pull of which we are 
measuring at some distance bigger than $1/\M$. 
If the SM massive particles are present, they create virtual 
particle-antiparticle pairs and ``polarize'' the vacuum around the source. 
The pair consists of a virtual positive energy state which is 
gravitationally  attracted to the source  
and a virtual negative energy state which is repelled from the 
source.  Therefore, the vacuum is polarized with virtual 
``gravitational  dipoles''. As a result, 
these dipoles screen the original gravitational interactions.
Thus, Standard Model particles ``shield'' sources (and themselves as well) 
from strong gravity. The heavier the particle, the  more effective  
is the shielding.  

In the next sections we consider the  higher dimensional framework
in which we will study the effects of quantum gravity modes on the 
observable 4D physics. The ``shielding'' phenomenon described above
plays the crucial role in our considerations. The main motivation
for going to higher dimensions, as was already mentioned
in Introduction, is that some brane-world scenarios can 
provide an opportunity to produce the hierarchy between 
the  scales $\M$ and $\msm$ dynamically.

\section{The Five Dimensional Framework} \label{theframework} 

In this section we shall 
set  the 5D framework which  allows to lower
the fundamental Planck scale $M_*$ much below the field theory cut-off of 
the SM. In fact, this is the model of Ref. \cite{dgp}
which we shall discuss briefly. 

Consider five-dimensional Minkowski space with a 
standard bulk gravitational action
\beq
S_{\rm bulk }~=~\int d^{(4+1)}X~\sqrt{|G|}~{\cal L}\left
(G_{AB}, ~{\cal R }_{ABCD}, ~\Phi\right )~,
\label{act1}
\eeq
where the capital Latin indexes run over $D=(4+1)$-dimensional space-time.
$G_{AB}$ denotes the metric of 5-dimensional
space, ${\cal R}_{ABCD}$ is the 5-dimensional
Riemann tensor and $\Phi$ collectively denotes other possible fields. 
We shall assume that there is a
3-brane in this space. Although the 3-brane can be realized as a soliton
of the corresponding field equations, at this point we shall
keep our discussion as general as possible, and will simply treat the  
3-brane as a hyper-surface that breaks five-dimensional 
translational invariance. We split the coordinates in 5-dimensions 
as follows: $ X^A~=~(x^{\mu},~y)~,$
where Greek indexes run over the four-dimensional  world-volume,
$\mu=0,1,2,3$~, and $y$ is the coordinate transverse to the brane.
In order to reduce our discussion to its main point the brane will 
be taken to have  zero width\footnote{In fact, we assume that 
the brane width is of the order of  $1/\msm \ll 1/\M$, 
see discussions below.}. In this approximation the brane 
classical worldvolume action takes 
the form: 
\beq
S_{\rm 3-brane}~=~-T 
~\int d^{4}x~\sqrt{|{g}|}~,
\label{brane0}
\eeq 
where $T$ stands for the brane tension and 
${g_{\mu\nu}}=\partial_\mu X^A\partial_\nu X^B G_{AB}$ 
denotes  the induced metric  on  the brane.
For simplicity of arguments we neglect brane fluctuations\footnote{
One could do this by, e.g., putting the brane onto an 
orbifold fixed point in extra dimension. In this case 
the brane is just an ``end'' of the infinite extra 
dimension.}. and go to the coordinate system
where the induced metric takes the following form:
\beq
{g}_{\mu\nu}~(x)~=~G_{\mu\nu}\left (x, y=0 \right )~.
\label{gind}
\eeq
We assume that the worldvolume theory on the brane is some gauge theory
that includes Standard Model and its possible high-energy extensions
(such as models of Grand Unification), with a field theory cutoff $\msm$.
The effective low-energy theory in the bulk is just 5D gravity with a
fundamental Planck scale $M_*$, above which quantum gravity effects become
important. The crucial assumption is that $M_*\ll \msm$, and in fact
we will be mostly interested in the case when $M_* \ll $ 1 TeV.

Before we proceed further let us make a digression and 
address the issue whether it is possible to have 
the scale of the worldvolume theory $\msm$ much bigger than 
the bulk fundamental scale $\M$ within any dynamically realizable
framework.  As an existence proof
of such a scenario we give  an  example of 4D
${\cal N}=1$ supersymmetric $SU(N)$ Yang-Mills theory
with large number of colors, $N\gg 1$. In this model
there are BPS domain walls \cite {GiaMisha} which in many respect 
resemble D2-branes of string theory \cite {Witten}. In particular,
the tension of this wall scales as $N\Lambda_{\rm SYM}^3$ 
\cite {GiaMisha} and  the width of the wall  scales as 
$1/ (N\Lambda_{\rm SYM})$ ($\Lambda_{\rm SYM}$ 
being the strong interaction scale) \cite {dgk} 
as it would for a D2-brane. 
The worldvolume theory of this  toy $(2+1)$ dimensional 
``braneworld'' provides a precise realization of the scenario 
which we are alluding to in the present work. 
Indeed, the bulk fundamental scale in this model is
$\Lambda_{\rm SYM}$ (the counterpart of our $\M$). 
However, there is much higher scale 
present in the theory, that is $ N\Lambda_{\rm SYM}\gg \Lambda_{\rm SYM}$ 
(the counterpart of our $\msm$).
In other words, there is the distance scale in the model, 
$1/N\Lambda_{\rm SYM}$, which is much smaller than
the fundamental length scale $1/\Lambda_{\rm SYM}$.
The brane width in this theory is determined by the 
shorter  scale $1/N\Lambda_{\rm SYM}$  and not by the fundamental 
scale $1/\Lambda_{\rm SYM}$ \cite {dgk}.
Furthermore, as was argued in Ref. \cite {GigaMisha}
there should exist in the theory nonperturbative states the mass
of which scale  as $N \Lambda_{\rm SYM}$. These states can be present 
as in the bulk as well as in the worldvolume theory. 
Moreover, the worldvolume states consist of the localized 
Goldstone particles and the heavy states the masses of 
which scale as $N \Lambda_{\rm SYM}$. 
Thus, the true ultraviolet cutoff of the worldvolume theory 
should be $N \Lambda_{\rm SYM}$, which is much bigger than the 
bulk fundamental scale $\Lambda_{\rm SYM}$ at which the bulk 
theory changes dramatically its regime 
due to the confinement.

We might hope that a similar scenario can be 
realized in string theory. Here the origin
of the huge scale separations between $\M$ and $\msm$ 
could be provided, for instance,  by a very small string 
coupling constant $g_s$.
The small string coupling gives rise to a new nonperturbative 
scale in string theory which is related to the fundamentals string length
as $g_s l_s$ \cite {Shenker}. 
The $g_s$ should play the role of the small number  
similar to that played by the parameter $1/N$ in 
the aforementioned example of supersymmetric gluodynamics. 
These issues will not be discussed in this paper, 
but will rather be postponed until 
further investigations. 

After this digression let us turn back to 
our main discussions. 
Thus, the unified tree-level bulk-brane action can 
be written as follows:
\beq\label{bulk-brane}
S~=~\M^3\int d^4x~dy ~\sqrt {|G|}~\left \{ {\cal R}~+~
{\cal O} \left ( { {\cal R}^2\over \M^2} \right ) \right \}
~+~\int d^4x ~\sqrt {|{g}|}~~{\cal L}_{\rm SM}(\Psi,~\msm)~.
\eeq
Here the first term is the standard 5D Einstein-Hilbert  
action, whereas the last  term describes coupling of bulk gravity to 
the brane world-volume SM field theory which has  the cutoff 
$\msm\gg \M$. 
We assume that the SM fields are confined to the 
brane.
Moreover, in what follows we will imply, without manifestly 
writing it that the Gibbons-Hawking surface term is included in the 
brane action as it gives rise to the correct bulk Einstein equations. 

What would be the observational consequences of the action 
(\ref{bulk-brane})?
Naively, the theory based on such an action is ruled 
out by everyday gravitational observations;  the 
extra dimension is not compactified, it has an infinite volume and
a brane-localized observer would measure the strong five-dimensional 
gravity with the small Planck scale $M_*$ 
rather than the weak 4D gravity  with
$M_{\rm P}\sim 10^{18}$ GeV.  However, the above naive argument 
is false; the reason being that  the important terms which 
are compatible with all the symmetries of the action have been 
left out in (\ref{bulk-brane}).  Such terms, even if not
included in the classical action, will be generated by quantum 
loops on the brane. To see this let us concentrate on the  
one-loop diagram of Fig.1.

\vspace{5mm}
\centerline{\epsfig{file=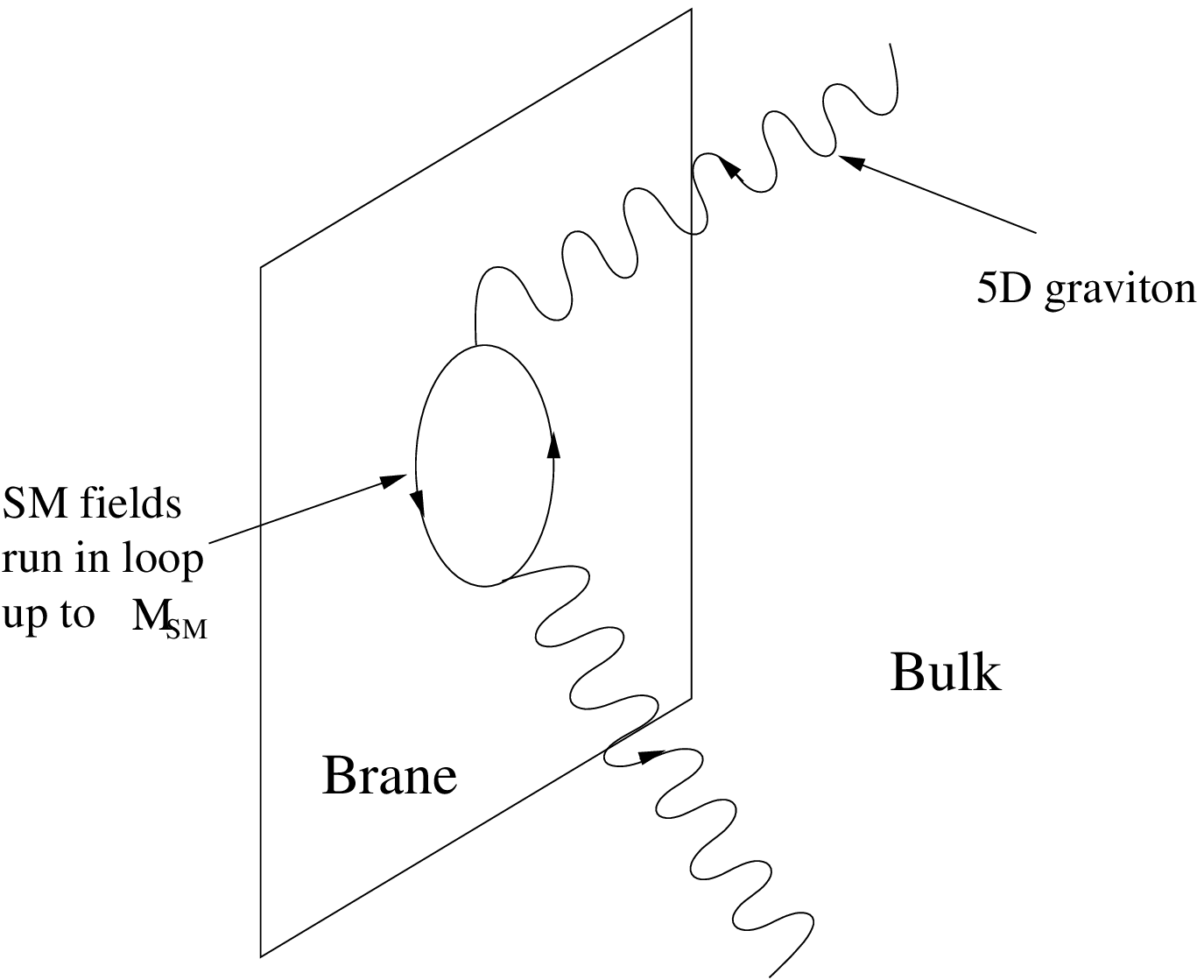,width=10cm}}
{\footnotesize\textbf{Figure 1:} SM fields on the brane
induce localized kinetic term for the bulk fields.}
\vspace{5mm}

This diagram
describes the renormalization of the graviton kinetic term, due to the
SM matter loops localized on the brane. Just as in the 4D case, the 
corresponding operator which is induced
by this correction has the  form \cite {dgp,DG} 
\beq
S_{\rm ind}~=~M_{\rm ind}^2\int d^{4}x~
\sqrt{|{g}|}~{\R4}(x)~,
\eeq
where $g$ is the higher dimensional metric evaluated at the position
of the brane defined in Eq. (\ref{gind}),  and ${\R4}(x)$ is the 
corresponding four dimensional Ricci scalar. 
As in the 4D case of the previous section 
the induced gravitational constant has to 
equal to the 4D Planck mass, $M_{\rm ind}^2 =\mpl^2$. 
This term should be added to the action (\ref{bulk-brane}).   

Thus, the bulk graviton acquires a four-dimensional 
kinetic term which is localized on the brane.
To realize the  importance of this correction, note that 
this term is weighted by the factor $\mpl^2$ 
which is much bigger than the bulk scale $\M$ that multiplies
the bulk Einstein-Hilbert term. As we shall see the scale $M_{\rm P}$ 
will play the role
of a 4D Planck scale for an observer on the brane. 
In this  framework, similar to  the 4D case, the  Planck scale is   
determined by  the cut-off of the Standard Model.
The high SM cutoff $\msm$ makes 
its own gravitational coupling to be naturally small. 
Thus, the  SM ``shields'' itself from strong 5D bulk gravity by 
means of the vacuum polarization effects described
in the previous section.
 
Let us remark here that 
in addition the following 4D induced terms should be  
included in the worldvolume action:
\beq
S_{\rm add}~=~{M_{\rm P}}^2~\int d^{4}x~
\sqrt{|{g}|}~\left [~ {\Lambda}+{\cal O}
\left ( {\R4}^2  \right ) \right ]~,
\label{4DR}
\eeq  
where $\Lambda$ in (\ref {4DR}) denotes  an 
induced four-dimensional cosmological 
constant. The role of this term is to renormalize 
the brane tension. In five-dimensional Minkowski space
a brane with nonzero tension inflates \cite{Vilenkin,Sikivie,kaloper}.
Therefore, to avoid  worldvolume inflation we fine-tune the 
brane tension $T$ and the brane worldvolume cosmological constant 
$\Lambda$ so that the net tension is vanishing
$T^\prime\equiv ~T-{\Lambda}~M_{\rm P}^2~=~0~.$
This is a usual fine-tuning of the four-dimensional 
cosmological constant.

\subsection{Four-dimensional Gravity on a Brane}

Here we recall how the 4D gravity is obtained on the 
brane with uncompactified infinite volume 
extra dimension (see Ref. \cite {dgp}).
At the end of the subsection we obtain certain new results
on the localization of massive fields on a brane.

We start by including 
in the action (\ref {bulk-brane})
the induced 4D Einstein-Hilbert term. Let us neglect 
for a moment the higher derivative terms 
(they will be discussed in the next subsection). 
The action takes the form:
\beq
S~=~\M^3~\int d^4x~dy ~\sqrt {|G|}~{\cal R}~+~
\int d^4x ~\sqrt {|{g}|} \left \{ \mpl^2~  ~R(x)
~+~{\cal L}_{\rm SM}(\Psi,~\msm)~\right \}~.
\label{ac}
\eeq
As before we imply the presence of the Gibbons-Hawking surface 
term on the worldvolume. The graviton propagator resulting from 
such an action is quite peculiar.
Ignoring the tensor structure for a moment we obtain the 
following expression for the two-point Green function \cite {dgp}:
\beq
{\tilde G}(p,y)~=~{1\over 2M_{*}^{3}~p~+~M_{\rm P}^2~p^2}~\exp\{-p|y|\}~.
\label{G5}
\eeq
Here $p^2$ is the  four-dimensional Euclidean momentum and
$p\equiv\sqrt{p^2}$. For 
sources which are localized on the  brane, i.e., for $ y = 0$, 
this propagator reduces to a massless 
four-dimensional Green's function
\beq
{\tilde G}(p, y=0)~=~{1\over \mpl^2~p^2}~+...,
\label{G5massless}
\eeq
provided that $p\gg 1/r_c\equiv M_{*}^3/M^2_{\rm P}$. Thus, at distances 
$r\ll r_c$  we observe  the correct
Newtonian behavior of the potential created by a static source of mass $M$:  
\beq
U( r\ll r_c )~= ~{M\over 16\pi M_{\rm P}^2~r}~+....
\eeq
At large distances, $r \gg r_c$, however, the behavior of the Green
function changes 
\beq  \label{G5massless1}
{\tilde G}(p, y=0)~=~{1\over 2\M^{3}~p}~+....
\eeq
This gives rise to the  Newtonian potential 
which scales in accordance with the laws of a five-dimensional theory
\beq
U ( r\gg r_c )~=~{M \over 16\pi^2~\M^3~r^2}~+....
\eeq
The explicit corrections to these expressions can be found in Ref. 
\cite {dgp}\footnote{The crossover behavior in this theory is similar to an 
otherwise very different model 
of\cite{rubakov}, in which gravity is also becoming five-dimensional at 
large scales. Note that the long-distance modification of gravity was 
suggested earlier in \cite {ian}  in a different context. None of these 
possibilities will be considered in the present work.}.
This somewhat unusual behavior can be understood in two equivalent ways
which we briefly discuss.

First let us adopt the  five-dimensional point of view.
In this language, although there is no localized massless particle, 
there exists a localized unstable  state in the spectrum (
we call it a resonance state for convenience).
The lifetime of this resonance is  $\sim 
r_c$. The resonance decays into the continuum of modes. 
This can be manifestly seen using the  
K\"allen-Lehmann representation for the Green's function
\beq
{\tilde G} (p,y=0)~=~{1\over \M^3}~ \left ( 
{1\over 2~p~+~r_c~p^2}\right )~ = ~
\int_0^\infty {\rho (s)~ds \over s~+~p^2 }~,
\label{KL}
\eeq
where the spectral density as a function of the Mandelstam variable $s$
takes the form:
\beq
\rho (s) ~\propto ~{1\over \sqrt{s}}~{r_c\over 4~+s~r_c^2}~.
\eeq
As $r_c\rightarrow \infty $ the spectral density 
tends to the Dirac function, $\rho(s) \rightarrow {\rm constant} ~\cdot~ 
\delta(s)$,
describing a stable massless graviton (this corresponds to the limit 
when the bulk kinetic term can be neglected).
At the distances $r\ll r_c$ the resonance mimics the massless 
exchange, and therefore  mediates the $1/r^2$ force. 
At larger distances, however, it 
decays into the continuum states and the 
force law becomes that of a five-dimensional theory, $\sim1/r^3$.

A different but equivalent way to 
understand  the above result is 
to adopt the point of view of the four-dimensional mode expansion.
The analysis of the linearized equation for the small fluctuations shows 
(see Appendix A) that there is a continuum of 4D massive states  with 
wave-function profiles  $\phi_m(y)$ which are
suppressed at the location of the brane by the following factor 
\begin{equation}
|\phi_m(y=0)|^2~ \propto ~{4\over 4~+ ~m^2~r_c^2}~,
\end{equation}
where $m$ denotes the continuous mass parameter of the Kaluza-Klein 
(KK) modes.
The Newtonian potential  on the brane is mediated by the exchange of all 
these Kaluza-Klein modes.  These give rise to the  expression:
\begin{equation}
U ( r )~ \propto ~{M\over M_{*}^3}~ ~\int_{0}^{\infty} ~{dm \over 4+ 
m^2r_c^2}
~{e^{-mr}\over r}~.
\end{equation}
At any distance $r$ the dominant contribution comes 
from the modes lighter than $m=1/r$.
The modes with $m < 1/r_c$ have 
unsuppressed wave-functions on the 
brane. Therefore, for  $r > r_c$ the  picture is similar 
to that of a  five-dimensional theory. 
In contrast, when  $r<r_c$ the picture changes 
since the modes with $m > 1/r_c$ have 
suppressed couplings on the brane. Although the  
number of the modes which  participate in the exchange
at a given distance $r < r_c$ is the same as 
in the five-dimensional picture, 
their contributions  are suppressed \cite{dgkn,Lykken}. 
Thus, the number of  the light modes 
effectively contributing to the exchange ``freezes-out'' 
and the resulting  behavior of the potential is $1/r$.

The same consideration can be applied  to other spin states,
for instance, to scalars \cite {dgp,DG,Lykken} or to  gauge fields 
\cite {dgshifman,archil,rubakov1}. In general, the picture is similar:
One obtains 4D behavior for $r < r_c$ and 5D behavior at  
$r > r_c$. 

We will investigate these properties further by
adding mass terms for the bulk fields. This is in particular 
important for scalars the mass terms of which 
are not protected by any symmetries.
Below, we will discuss a scalar field which 
has a nonzero mass terms in the bulk and on the brane
(the same consideration applies to other 
massive fields as well).
Neglecting all other fields the action takes the form:
\beqa 
S~=~M_*^3 \int {\mathrm d}^5 X~ 
\Big \{  \left [\de_A \Phi \right ]^2~ -~ M_B^2 ~\Phi^2 \Big \}~+~
M_{\rm P}^2 \int {\mathrm d}^4 x 
\Big \{ ~\left [\de_\mu \Phi(x,~y=0)\right ]^2~-~\mu^2 ~\Phi^2 \Big \}~.
\label{scal1}
\eeqa
We choose somewhat unconventional normalizations where 
the scalar field is dimensionless. 
This system is analyzed in detail in  Appendix 
along the lines of what we did for the massless case. 
The resulting  propagator has the form 
(in Euclidean space) 
\beq 
{\tilde G}_(p,y)~=~{\exp\{-\sqrt{p^2~+~M_B^2} ~|y|\} 
\over 2M_{*}^{3}~\sqrt{p^2+M_B^2} ~+~M_{\rm P}^2~(p^2+\mu^2) }~.
\label{G52}
\eeq
For $p \gg 1/r_c$ and at $y=0$, the propagator 
resembles that of  a four-dimensional field of mass $\mu$, i.e.,  
${\tilde G}(p,0) \sim (p^2 + \mu^2)^{-1}$. As before, 
it is the four-dimensional part of the action that determines the
short distance behavior. 

Moreover, as in the massless case,
we could study the four-dimensional mode expansion. 
The detailed analysis (see Appendix) 
reveals that we should distinguish two cases.  
We will discuss them in turn.

$\{1\}$  For $\mu > M_B$ there is no zero mode, 
but rather  a continuum of massive modes with mass 
$m \in [M_B, \infty )$. The wave-functions of the continuum modes
have the following transverse space  
profiles at the position of the brane:
\beq\label{profile}
|\phi_{m}(y=0)|^2 \propto \left [ 4~+~r_{c}^2~m^2~
\left \{ \frac{(1~-~\mu^2/m^2)^2}{1~-~M_B^2/m^2} \right \}  
\right ]^{-1}~.
\eeq  
As shown in Appendix, this profile results in the 
suppression of all  the continuum modes on
the brane  except those in a narrow  mass 
band of the width $\sim 1/r_c$ centered 
around the value of $\mu$. In other words, 
to a four-dimensional observer the  continuum
of modes effectively appears as  a single meta-stable 
mode of the mass $\mu$.

$\{2\}$ For $\mu < M_B$  we still have a continuum of massive modes starting
at $m=M_B$  with the same profiles at the brane position given by  
Eq. (\ref{profile}).
In addition we also find a 
\emph{normalizable mode} of mass $\sim \mu$. 
This is to be interpreted as a truly four-dimensional 
localized state with the well defined 4D mass. 

The existence of this localized mode 
can also be seen from the propagator (\ref{G52})
which has a  \emph{physical} pole at $p^2 \sim \mu^2$ if $\mu<M_B$. 
Since the value $m\sim \mu$ is outside the continuum band 
in this case all the 
continuum modes are suppressed on the brane. Therefore,
a four dimensional observer will still effectively see 
a single state of mass $\sim \mu$,  but this
time a true 4D localized state. 

It is very interesting to note that for $\mu =0$ 
(i.e., no mass term on the brane) 
but nonzero $M_B$, still   there is  a bound state with the mass 
$m^2_{BS} \sim {M_B}/{r_c}$ which is localized on the brane.
This mass is  very small in the regimes under consideration. 
Therefore, this 
framework provides  a new mechanism for the localization 
of an almost-massless particles on a brane in an infinite
volume flat extra dimension.

\subsection{Tensorial Structure of the Propagator}

We have seen in the previous section that the usual 4-dimensional
Newton  law for gravity is reproduced at distances $r \ll r_c$.
At very short distances, however, we expect  
the Newton law to be modified by higher-derivative terms 
which we did not consider so far. 
Moreover, we neglected in the previous subsection
the tensorial structure of the Green function, however,
the predictions for the 
relativistic effects strongly depend
on this structure. In this subsection we will address these 
issues. 

In our model, it is not immediately obvious  
which scale determines the modification due to the quantum gravity
corrections. This question for scalar field theory models 
was studied in Ref. \cite {adgp}
where it was concluded that the modification occurs 
for distances of order  $1/\M$. 
Here we investigate this issue 
for the gravitational action and in 
addition  study an important  question 
of the tensorial structure of the graviton  propagator.

Since the field theory of gravity is non-renormalizable 
it  should be regarded as a low-energy effective theory. 
The effects of quantum gravity at low energies can be 
encoded by adding  all possible higher-derivative 
operators to the gravitational action. 
In the bulk, gravity becomes strong at the scale $M_*$, 
hence, the higher-dimensional 
operators in the bulk are suppressed by powers of $M_*$.
We would like to study the effects of these 
terms on the propagator. For calculational convenience we 
choose the following form of the higher-derivative terms in the bulk:
\beqa
S_{\rm bulk}~ =~ \int d^5 X \sqrt{|G|} \, M_*^3~ \left({\cal R} + \, {c \over
M_*^2}~\left (  {{\cal R}^2\over 2}~-~{\cal R}_{AB} {\cal R}^{AB}\right 
)~+...\right),
\label{ba1}
\eeqa
where $c$ is some constant and 
dots denote all other possible higher-order operators
\footnote{The truncation of the action (\ref {ba1}) at any finite order
in derivatives  generically gives rise to  
unphysical poles in the propagator for the momenta of order $\M$
(unless the coefficients of the higher derivative terms are chosen
very carefully as in the Gauss-Bonnet term for instance). 
However, the expansion in powers of  
$p^2/\M^2$ breaks down in that domain so
these poles are spurious  and should be neglected.
In the total action, if it comes from a consistent
higher-dimensional theory, such as string theory,
there should not be any unphysical states.}.
As we discussed before, we also expect that 
additional higher-dimensional operators  will be induced on the brane, 
in analogy to the induced brane Ricci term.  The
strength of these operators, however, is suppressed by $\msm$. 
Below we will compute the modification of the 4D Green function 
on the brane due to the higher-order operators in the bulk 
action (\ref {ba1})
and discuss the corresponding tensorial structures.
 
We need to calculate the gravitational 
perturbations created by a static source which is 
localized on the brane. 
Let us  introduce the metric fluctuations:
\beqa
G_{AB} ~=~ \eta_{AB} ~+~h_{AB}~.
\eeqa
We choose {\it harmonic~ gauge} in the bulk: 
\beqa
\partial^A h_{AB}~ =~{1\over 2}~ \partial_B h^C_C~.
\label{gauge}
\eeqa 
The $\{\mu 5\}$ components of the  
equations of motion lead to the condition:
\beqa
h_{\mu 5} ~=~0~. 
\label{mu5}
\eeqa
Thus, the surviving components of $h_{AB}$ are 
$h_{\mu\nu}$ and $h_{55}$.
In harmonic gauge the $\{55\}$ component  of Einstein's equation 
can be solved by the substitution:
\beqa
~ \partial_A\partial^A ~h^5_5 ~=~
~\partial_A\partial^A~ h^\mu_\mu~.
\label{mn}
\eeqa
The indices  in all these equations 
are raised and lowered  by a flat space  metric tensor.
Finally, we come to the $\{\mu\nu \}$ components
of the Einstein equation. After some 
rearrangements  these    take  the form:
\beqa
\M^3~\left ( \partial_A\partial^A~h_{\mu\nu}
-{1\over 2} \eta_{\mu\nu} \partial_A\partial^A h^\alpha_\alpha
-{1\over 2} \eta_{\mu\nu} \partial_A\partial^A h^5_5 \right ) 
\nonumber \\
+\M^3~\left [{c\over \M^2} \left( -\partial_A\partial^A 
\partial_B\partial^B~h_{\mu\nu}
+{1\over 2}\eta_{\mu\nu} \partial_A\partial^A \partial_B\partial^B~
h^\alpha_\alpha
+{1\over 2} \eta_{\mu\nu} \partial_A\partial^A \partial_B\partial^B~h^5_5 
   \right )
  \right ]
\nonumber \\
+\mpl^2~\d~ \left ( 
\partial_\alpha \partial^\alpha~h_{\mu\nu}
-{1\over 2} \eta_{\mu\nu} \partial_\beta\partial^\beta h^\alpha_\alpha
+{1\over 2} \eta_{\mu\nu} \partial_\beta\partial^\beta h^5_5
-\partial_\mu \partial_\nu h^5_5
\right )~=~T_{\mu\nu}(x) ~\d~.
\label{basic}
\eeqa
Here, we choose such a  normalization 
that the energy-momentum tensor of a source   
localized on the  brane  is  $T_{\mu\nu}(x) ~\d~$.
Multiplying  both sides of these equations by $\eta_{\mu\nu}$
we obtain:
\beqa
\partial_A\partial^A~h^\alpha_\alpha~=~-{T^\alpha_\alpha~
\delta(y) \over 3M_{*}^{3}}~+~{c\over \M^2}~
\partial_A\partial^A \partial_B\partial^B~h^\alpha_\alpha~.
\label{trace}
\eeqa
Finally, using this expression we find:
\beqa
M_*^3\left (  
\partial_A\partial^A~-~{c\over \M^2}\partial_A\partial^A 
\partial_B\partial^B 
\right )~h_{\mu\nu}~+~
\mpl^2~\d \left (\partial_\alpha\partial^\alpha h_{\mu\nu} 
- \partial_\mu\partial_\nu h^5_{5}  \right)~=\nonumber \\
~\left( T_{\mu\nu}~ -~
\eta_{\mu\nu}{T^\alpha_\alpha\over 3}\right)\delta(y)~.
\label{eqn1}
\eeqa
Turning to Euclidean momentum space and multiplying both sides of the 
equation by a (probe) conserved energy-momentum tensor we find:
\beqa
\left( M_*^3 \left( (p^2-\pr_y^2) \,+\, {c \over M_*^2} (p^2 - 
\pr_y^2)^2 \right)\,+\, \mpl^2\,p^2\,\delta(y)\,\right) 
{\tilde h}_{\mu\nu}(p, y)~{\tilde T}^{\prime \mu\nu} = \nonumber \\
~\left \{  
{\tilde T}_{\mu\nu}{\tilde T}^{\prime \mu\nu} ~-~
{1\over 3}~{\tilde T}^\alpha_\alpha 
{\tilde T}^{\prime \beta}_\beta
\right \}~\delta(y).
\label{prop2}
\eeqa
Following \cite {DG} we look for a solution of this 
equation in  the following form
\beqa
{\tilde h}_{\mu\nu}(p, y)~{\tilde T}^{\prime \mu\nu}\,=\,A(p)~B(p,y)~,
\label{form}
\eeqa
where the function $B$ satisfies the equation:
\beqa
\left( (p^2-\pr_y^2) \,+\, {c \over M_*^2} (p^2 - 
\pr_y^2)^2 \right)~ B(p,y)~ = ~\delta(y)~.
\label{fung}
\eeqa
The expression for the propagator on the brane
is as follows:
\beqa
{\tilde h}_{\mu\nu}(p, y=0)~{\tilde T}^{\prime \mu\nu} \,=\,
{{\tilde T}_{\mu\nu}{\tilde T}^{\prime \mu\nu} ~-~
{1\over 3}~{\tilde T}^\alpha_\alpha 
{\tilde T}^{\prime \beta}_\beta   \over 
\mpl^2 \, p^2 \,+\, M_*^3\, B^{-1}(p,0)}\,.
\eeqa
Furthermore, 
we can calculate  $B(p,0)$ from (\ref {fung}):
\beqa
{B^{-1}(p,0)}\,\simeq\, 2\,p\left ( 
1 +{\sqrt{c} \,p\over M_*} + \ldots  \right)\,.
\eeqa
Using this expression we find
the propagator between two points on the brane\footnote{The scalar 
part of this propagator was obtained in \cite {adgp}.}:
\beqa
{\tilde h}_{\mu\nu}(p, y=0)~{\tilde T}^{\prime \mu\nu}
\,=\,{{\tilde T}_{\mu\nu}{\tilde T}^{\prime \mu\nu} ~-~
{1\over 3}~{\tilde T}^\alpha_\alpha 
{\tilde T}^{\prime \beta}_\beta 
\over \mpl^2 \, p^2 \,+\, 2\,M_*^3\,p\, 
[1 +\sqrt{c}~ p~M^{-1}_* + \ldots ]}\,,
\label{prop3}
\eeqa
where the dots denote terms which are of higher order in $p/M_*$.
We assume that gravity above this scale becomes soft.
As was emphasized in Ref. \cite {dgkn}, and will be shown below
in detail, there exists in our model a well-defined expansion 
of SM scattering cross-sections and other SM 
observables  in powers of the usual four-dimensional Planck
mass $\mpl$. In the leading order in this expansion 
the usual Standard Model
results are reproduced for any energy scale. 
The gravitational corrections  for energies below $\M$
can be calculated within the standard framework. 
However, at higher energies the effective gravitational action
ceases to be valid and the quantum gravity corrections should  
be taken into  account. Since we assume the   
``softness'' of quantum gravity effects these corrections 
should remain negligible compared to the SM corrections.

As we pointed out before, the bulk quantum 
gravity scale $\M$ can be smaller that 1 TeV.
At distanced below $1/\M$ the Newton law is modified.
This  law has only be tested at  distances 
bigger that  0.2 mm \cite{measurements}. Therefore,  
a model with $\M \ge 10^{-3}$ eV does not contradict the 
data on static force measurements.  
Note that for such low values of $M_*$, the cross-over to 
five-dimensional gravity only occurs for $r>10^{63}$ cm
\cite {dgkn}.

We would like now to discuss the 
tensorial  structure of the graviton propagator in the 
present model. In Eq. (\ref{prop3}), 
the tensorial structure is similar to that of
a 5D graviton (or equivalently of a 4D massive graviton) 
\cite{dgp}. This points to the discontinuity which leads to the 
contradictions with observations \cite{Veltman,Zakharov}. 
However, this problem 
is an artifact of using the lowest tree-level 
approximation \cite {Vainshtein}.  
We discuss  below two ways to avoid this problem.

In the context of infinite volume 
uncompactified extra dimension 
we note that the lowest tree-level approximation which was used 
to derive  (\ref{prop3}) breaks down at small distances 
\cite {Vainshtein}, \cite {Arkady}. 
In fact, the tensorial structure obtained in (\ref{prop3}) 
is applicable  for distances $r \gg r_c$ where the 5D behavior takes over.
For short distances $r\ll r_c$ 
the higher corrections become dominant. Thus, 
one has to sum up all the tree-level graphs which are obtained 
by iterations of the nonlinear Einstein equations 
in the external source. This is equivalent of 
finding exact solutions to the classical equations of motion.
The net result of this, as was advocated in Ref. \cite {Arkady},
is that the coefficient $1/3$ in the numerator of (\ref{prop3}) 
is promoted to a momentum dependent formfactor. 
For small momenta (i.e., large distances $r\gg r_c$) 
the formfactor turns into the coefficient
$1/3$, however, for large momenta, i.e., small distances 
it returns the value $1/2$, consistent with the 4D observations. 

The Schwarzschild solution in this case can only be found in the 
approximation $r_c\to \infty$ \cite {Dick,Arkady} which 
by construction has no discontinuity. Moreover, some other 
exact cosmological solutions were  found \cite {cedric,lu,Dick}  
which demonstrate that
there is no discontinuity in the full classical theory \cite {Arkady}.
Hence, the extra helicity $\pm 1$ and 0 states 
of the 5D graviton decouple  from the 4D matter fields as $r_c\to 
\infty$.

Note that the continuity in the graviton mass in curved 
(A)dS backgrounds was demonstrated recently in 
Refs. \cite{Kogan,Massimo} (see 
also further considerations in Ref. \cite {Kogan1}). We should emphasize 
that Refs. \cite{Kogan,Massimo} as well as 
our works discuss the continuity in the classical 
4D gravitational interactions with 4D matter.
There certainly is the discontinuity in the full theory in the  sense that 
there are extra degrees of freedom in the model. The latter
can manifest themselves at the quantum level in loop diagrams 
\cite{Duff}. However, what is important for observations
is the continuity in the tree-level couplings of gravity to matter.
These couplings are continuous.

In general, the simplest  
way to  deal with the discontinuity problem, 
as was suggested in Ref.
\cite {dgkn}, is to compactify the extra space
on a circle of a huge radius $R$ . This radius 
can be bigger that the present day horizon distance, 
but still smaller that $r_c$. 
For instance, if $\M \sim 10^{-3}~{\rm eV}$, then 
$r_c \sim 10^{63}~{\rm cm}$ and $R$ can be as large as $10^{59}~{\rm cm}$
\cite {dgkn}. 
This is about thirty orders of magnitude bigger that the 
horizon scale; thus, this  extra dimension is infinite for any 
practical purposes.

The convenience of such a procedure is that 
in this case the lowest tree-level approximation
to the graviton exchange becomes applicable even at distances $r\ll r_c$
(so there is no need to sum up all the tree level graphs).
The reason is that there is a zero mode  
which gives the correct 4D coefficient $1/2$ in the tensorial
structure, and moreover, all the KK modes which could, in the conventional 
case, turn this coefficient into $1/3$ are now additionally 
suppressed by the ratio $R/r_c$.
This is possible to see from the 4D expression for the 5D graviton 
propagator \cite {dgkn}: 
\beqa\label{novdvz}
G_4^{\mu\nu;\alpha\beta}(p)~ \simeq
~\left ( {1\over
2}(\eta^{\mu\alpha}\eta^{\nu\beta}~ +~ \eta^{\mu\beta}\eta^{\nu\alpha})~ -~ 
{1\over
2}\eta^{\mu\nu}\eta^{\alpha\beta}  \right )~{1\over p^2} ~+ ~\nn \\
 \frac{1}{\pi^2} ~\frac{R}{r_c}~
  \sum_{n=1}^{\infty}~{1 \over n^2}
 \left ( {1\over
2}({\tilde \eta}^{\mu\alpha}{\tilde \eta}^{\nu\beta}~ + ~
{\tilde \eta}^{\mu\beta}{\tilde \eta}^{\nu\alpha})~ - 
~{1\over
3}{\tilde \eta}^{\mu\nu}{\tilde \eta}^{\alpha\beta} \right )
~{1\over p^2~+~m_n^2~}~ ,
\label{sumG}
\eeqa
where 
\beq
{\tilde \eta}_{\mu\nu} ~=~ \eta_{\mu\nu}~+~{p_\mu~p_\nu\over p^2}~.
\eeq
The first terms in this expression corresponds to the 
4D massless mode and the rest of the terms correspond  to the 
KK modes which due to the induced kinetic term on the brane are
suppressed by $R/r_c$ \cite {dgkn}. Thus, the model has no  
discontinuity even in the lowest order tree-level approximation.

\section{Gravity above $M_*$}

In this section we will try to summarize   
the main qualitative reasons why our framework survives 
all the constraints. 
As we shall see, the reason is the self-shielding of the SM 
fields from the bulk theory. 
The SM generates a large 
brane kinetic term for any bulk field coupled to the SM particles
and makes it  to be weakly coupled to the brane matter.
In other words, the high-dimensional strongly 
coupled bulk theory is ``projected'' onto
a more weakly  coupled four-dimensional 
counterpart on the brane.
This projection, as we will see below, 
only takes place in an intermediate range of energies
$\sqrt{M_PM_*}\gg  E \gg  1/r_c$  
above which brane-bulk interactions become strong.
However, this window is 
large enough to be compatible with all the existing
data. We shall present 
the discussion in two steps. Fist we will give a purely field
theoretical  consideration, without referring to the 
precise nature of quantum gravity above the scale $M_*$;
then, in the next section, we will assume  
that the bulk gravitational theory above $\M$ 
possesses some  generic properties of string theory, 
and show that in this case all the experimental 
constraints can be satisfied.

What is the lower bound on the scale $M_*$? 
It is impossible to answer this 
question without making 
assumptions about the nature of quantum gravity 
above this scale. However, the  following 
general considerations  should be valid.   
The usual formulation of general relativity is appropriate up to scales 
of the order of the fundamental Planck scale $M_*$. So  we 
must think  of  GR  as an effective field theory, valid at energy 
smaller than $M_*$; moreover, we expect it to be embedded into a more 
fundamental theory that regulates the ultraviolet behavior. 
Whatever this theory is, it is reasonable to assume that its 
effect is to make quantum gravity ``softer''  at energies above
$M_*$,  i.e.,  to regularize 
the strength of graviton  self-interactions
and that of the interactions  of gravity with matter.
The fact that this should be the case is suggested by the only known 
consistent theory of quantum gravity, that is string theory.
This theory exhibits a well known softening of scattering amplitudes 
at high momenta.

As an example of the soft behavior one could consider (see Appendix) 
the interaction potential between two static sources  
in string theory. This potential has no  
short-distance singularity, as opposed to the case of 
a static potential obtained in field theory
which is singular at the origin.
In a field theoretical computation one could in principle 
also get such a  smooth result if the
propagator of the intermediate virtual state  
vanishes faster than $1/p^2$ in the ultraviolet, i.e.,  
$ p \to \infty$. This could effectively be 
described  by introducing a certain form-factor
$f(p)$  in the graviton propagator (and/or in vertices) 
such that $f(p) \to 1 $ for small $p$ and $f(p) \to 0 $ for $p$ larger 
than  $M_*$. This would have the effect of cutting-off  the momentum 
in the graviton internal lines of  any Feynman diagram above $M_*$.

As it was shown in \cite{dgkn}, under these circumstances
the high-energy  colliders production processes 
of particles or the process of  star cooling  place essentially 
no constraint on the scale $M_*$. 
The reason for this is the ``self-shielding'' of the brane-localized 
Standard Model from the strong bulk gravity. This manifests itself in the 
aforementioned  
suppression of the heavy KK wave functions on the brane. 
As a result, 
their production in any high-energy process on the brane is 
extremely suppressed. 
For instance, consider the rate of the bulk graviton 
production in a SM process at 
energy $E$. The rate is given by
\begin{equation}
 \Gamma ~\sim ~{E^3 \over M_*^3}~\int_0^{m_{max}} ~|\phi_m(0)|^2 ~dm~,
\end{equation}
where the integration 
is  over the continuum of KK states  up to the maximum 
possible mass which  can be 
produced in a given process, i.e., $m_{max} \sim E$. 
Since the wave-functions of the heavier KK states  
are suppressed on the brane by a  factor ${1/ 
m^2r_c^2}$ the integration domain is 
effectively truncated at $m \sim 1/r_c$. Thus, one obtains  
\begin{equation}
\Gamma ~\sim~ {E^3 \over M_*^3~r_c} ~\sim ~{E^3 \over M_{\rm P}^2}~.
\label{rt}
\end{equation}
If we were to neglect the induced kinetic term on the brane
the rate would be given instead by the ratio 
$E^4/M_*^3$ which is  unacceptably large. 
On the other hand, the rate  (\ref {rt})
is of the order of the  production rate for  
a single four-dimensional massless graviton 
and is totally negligible.
Although  gravity ``becomes strong'' at the scale $M_*$, 
the gravitational loop  corrections to any Standard Model 
amplitude would be absolutely negligible  even though  
the  momenta in the internal  lines are above $M_*$.

\vspace{5mm}
\centerline{\epsfig{file=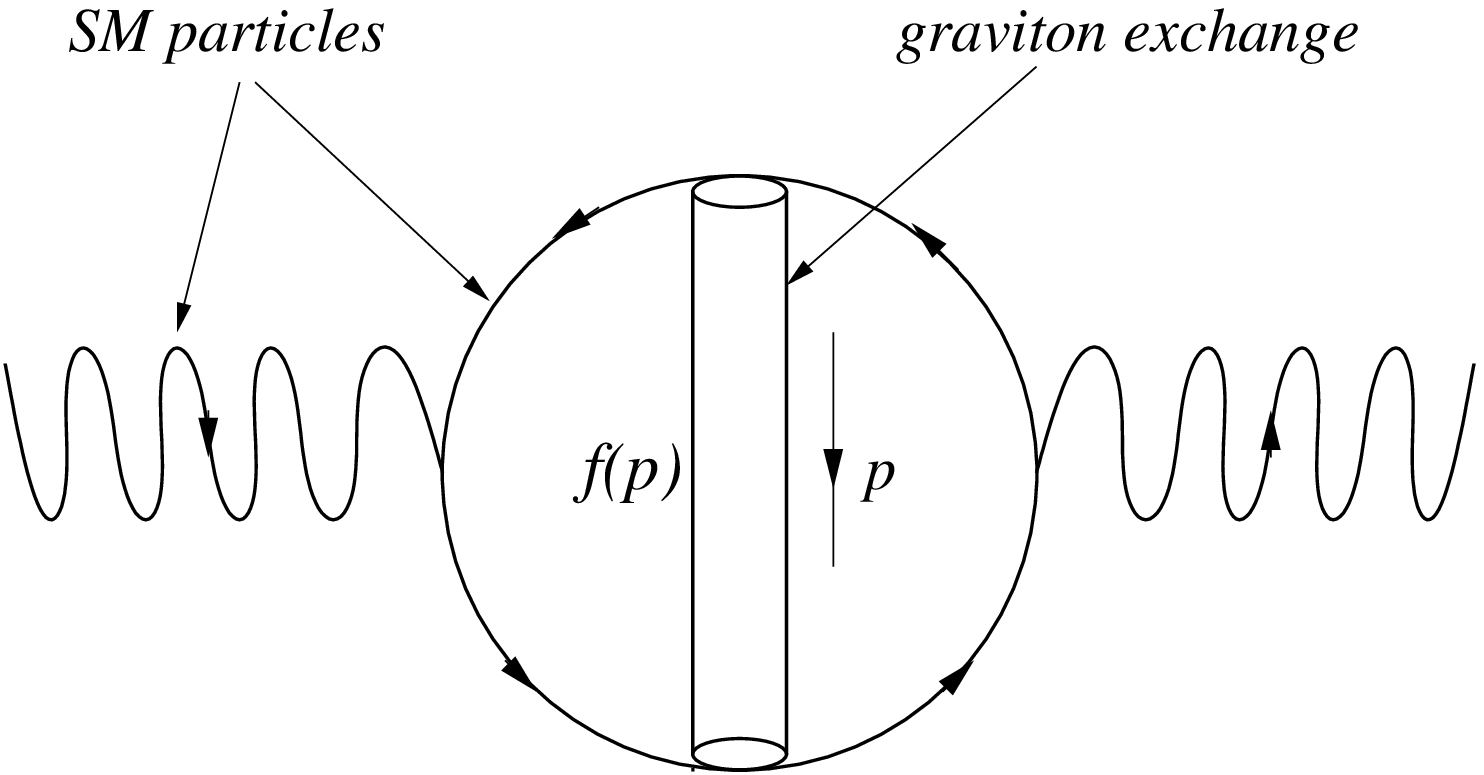,width=10cm}}
{\footnotesize\textbf{Figure 2:} Graviton loop correction to a standard 
model amplitude.
The 'tube' represents a graviton propagator above the energy $M_*$. The 
form factor $f(p)$ 
effectively cuts off the graviton momentum $p$ at $M_*$.  }
\vspace{5mm}

Consider for example the diagram in Fig.2. 
The form factor effectively switches 
off the graviton propagator (represented
by a ``tube'' in Fig.2 )  when $p >  M_*$. Thus,  
the diagram is dominated by the momentum running
in the matter lines which could be as large as  $\msm \gg M_*$. 
Due to the smallness of the matter-gravity coupling
this diagram will give a correction which is suppressed  
compared to the one due to the gauge boson replacing 
the graviton line.    

In this respect we would like to point out one more advantage of the 
present framework. It deals with the gauge coupling unification.
We think of the scenario where  
the SM worldvolume scale is huge, 
let us say of the order of the GUT scale. 
In such a case the gauge coupling unification
is not affected by strong gravity corrections 
precisely because of the reasons 
outlined above. Thus, the prediction of 4D theory on the 
gauge coupling unification in supersymmetric models \cite {UNIFICATION}
will remain intact in this framework.

As we have shown before,
the only constraint comes from the measurement of 
the Newton force, which implies $M_* > 10^{-3}$ eV. 
This can be understood by using yet another language.
Consider the  Newton interaction between two static
sources. Without the cutoff the 
Newton  potential between two masses $m_1$ and $m_2$ 
takes the form:
\begin{equation}
V( r )~ =~-~{m_1~m_2 \over M_*^3}~
\int {d^3~\vec{p}\over (2\pi)^3}~ {\exp(~i~\vec{p}\cdot \vec{r}~)
 \over 2p ~+ ~r_c~{p}^2}~,
\end{equation}
which for $r\ll r_c$ reduces to the conventional  4D potential
\begin{equation}
V( r )~ = ~-~{m_1~m_2 \over M_{\rm P}^2}~\int 
{d^3~\vec{p}\over (2\pi)^3 }~{\exp(~i~\vec{p}\cdot 
\vec{r}~)
 \over p^2}~. 
\end{equation}
In order to introduce the effective 
cutoff, however, one has to include the form-factor:
\begin{equation}
V( r )~ = ~-~{m_1~m_2 \over M_{\rm P}^2}~\int 
{d^3~\vec{p}\over (2\pi)^3}~
{\exp(~i~\vec{p}\cdot
\vec{r}~)
~f(p) \over p^2}~, 
\end{equation}
where $f(p)$ dies off for the momenta above $M_*$. 
Such a theory would predict deviations from the ordinary Newton  law at 
distances $r< 1/M_*$.  This is because in the conventional case the  
gravitons with momenta $\sim 
1/r$ contribute at the distance $r$, 
whereas in the present case the contribution 
of such gravitons  is  suppressed by the formfactor $f(p)$. 
Since the gravitational 
law has been tested down to sub-millimeter  distances, we obtain 
the bound on $M_*$ from these considerations, 
$\M~\ge~(0.2 ~{\rm mm})^{-1}$.

\section{Modeling Quantum Gravity by String Spectrum}

Although we expect that the theory of quantum gravity should make graviton 
amplitudes ``softer'', nevertheless,
it is unlikely that the only effect of 
quantum gravity can be summarized in a form-factor that switches off 
graviton exchange at energies above $M_*$. 
In particular, quantum gravity could soften its amplitudes 
by introducing an  enormous multiplicity of states above $M_*$. 
This expectation is certainly supported by string 
theory which is at present the only candidate for 
a self-consistent theory of quantum gravity. 
String theory predicts an exponentially increasing number of 
states which  can be excited at energies above the string scale.
One of the implications of this fact is the Hagedorn phenomenon.

Therefore, if the theory of quantum gravity above $M_*$ is some version of a 
string theory, we have to face the existence of an exponentially large 
multiplicity of bulk states with the Regge recurrences 
governed  by the scale $M_*$. Naively, this 
ruins any hope of bringing the quantum gravity scale below 1 TeV. 
Indeed, it seems natural that if such a multiplicity of states 
is coupled to the SM particles  there is  no way for 
them not to manifest themselves in all possible high-energy processes. 
For instance, to exclude $M_* < $ KeV it would be enough to 
think of the interior of the Sun
where there is no sign of any exotic Hagedorn type behavior. 

The purpose of  this section is to show that this in fact is 
not the case. The absence of the Hagedorn catastrophe in
the scenario with  high-cut-off Standard Model 
can be  completely compatible both with very low 
value of $M_*$ as well as with an  exponentially increasing density 
of states. The 
detailed discussion of various bounds will be given in the following  
subsections. Here we shall summarize the main reasons why 
the present framework is not excluded. 

To be concrete we shall assume  that the bulk spectrum is that of 
a closed bosonic string theory (ignoring the tachyon). The main 
point is to find out what is the impact of the
Regge states on the SM which is localized on the brane. 
To answer this question we shall use some results that will be derived in 
great  detail in the next 3 subsections. 
Here we shall quote them  without a proof.

Consider a bulk field $A$ of some high integer spin which has a 
five-dimensional mass $M_B$ and a brane mass $\mu \sim M_B$.
In addition, it has both the  brane and bulk kinetic terms, 
just like our graviton. The Lagrangian of 
this field can be schematically written as follows:
\begin{equation}
M_*^3 \Big \{  \left [\de_{(5)} A \right ]^2~ -~ M_B^2 ~A^2 \Big \}~+~
\pl^2 ~\delta(y)~ 
\Big \{ ~\left [\de_{(4)}  A(x,~y=0)\right ]^2~-~\mu^2 ~A^2 \Big \}~,
\label{A}
\end{equation}
where $\pl$ is some scale to be specified below. In these notations
$A$ is dimensionless. Assume that the field $A$ couples derivatively to 
the localized matter on the brane and the corresponding 
coupling is defined by inverse powers of $M_*$. 
As it will be shown, in all the 4D processes the effect of 
such a bulk field is reduced to that  of  a single 4D 
state (of the same spin) which has the mass $\sim \mu$,
and the coupling square  proportional to  ${1/ \pl^2}$.
Depending on the spin of the state $A$, this has to be multiplied by an 
appropriate power of $p/M_*$  arising  from the derivatives in the original 
coupling. The crucial point is that the scale $\pl$ 
which is induced by the localized matter loops, 
depends on the number of derivatives in the coupling of the bulk 
field with the SM.  For a  field $A$ 
which is coupled with $n-1$ derivatives to the SM fermions 
the scale  is $\pl^2 \sim {M_{\rm P}^{2(n-1)} / M_*^{2(n-2)}}$. 
As a result, the coupling of this bulk field to the 
localized SM states takes the form:
\begin{equation}
{\rm Effective~coupling}~ \sim ~{p^{2(n-1)}~\over \mpl ^{2(n-1)}}~.
\label{efcoup}
\end{equation}
This is the very same mechanism 
by  which the Standard Model weakens its coupling 
to  the strong bulk gravity. 
Moreover, we note that the  higher is the power of the 
derivative interaction  the  larger is the induced 
4D kinetic term on the brane, and, consequently,
the weaker is the coupling of $A$ to the SM. 

Therefore, the  SM shields itself not only from strong bulk 
gravity but also from other bulk fields. Furthermore,
the higher is the spin of the bulk state the  more 
efficient is the shielding.  
Given this fact, it is easy to 
understand how the Hagedorn 
catastrophe is avoided:
it is true that the number of modes available at higher 
energies grows  exponentially, however, 
most of them are coupled with higher derivatives to the SM
and thus their effective 4D coupling becomes very weak 
(\ref{efcoup}). 

These  arguments show that the dominant 
contribution comes from spin-2 states that couple via a
single derivative to the  
SM fermions (note that there are also 
spin-2 states that couple via higher derivatives 
and therefore  are less important). Since the scale $\pl$
depends only on the number of derivatives in the coupling, 
then they all couple to the SM 
by the $M_{\rm P}$-suppressed interactions, just 
like an ordinary graviton. 
Therefore, their emission 
rate in high energy processes scales as follows:
\begin{equation}\label{rate} 
 \Gamma ~\sim ~{E^3 \over M_{\rm P}^2}~ n_{max}~.
\end{equation}
Due to the assumed Regge behavior   $n_{max} = (E/M_*)^2$. As a result,
Eq. (\ref{rate}) constrains $M_*$ to be above $10^{-3}$ eV.  
Remarkably, this bound coincides 
with that coming from sub-millimeter gravity measurements.

\subsection{String Spectrum} \label{string}

In the following  we consider a field theoretical model the 
spectrum of which mimics that of closed bosonic string theory
in critical dimension (neglecting tachyons). We will show that under 
certain assumptions  the enormous multiplicity of states accessible at  
low energies can be totally compatible with observation. 
Before doing so, we will briefly recall what 
the main features of the  string spectrum are. 

A generic closed  string state can be described by
two copies (left and right moving) of  an infinite set  of 
creation operators\footnote{   
One usually does this construction in the light-cone gauge where
all the obtained stringy states are physical \cite {GSW}. 
Since we are dealing only with the kinematical 
features of the string spectrum, we will not discriminate 
between the light cone-gauge construction and that 
of the Lorentz covariant formalism.}
$\alpha^{\mu\dagger}_n$, 
$\tilde {\alpha}^{\mu\dagger }_n$, where 
$\mu $ is a Lorentz index and $n$ labels the ``oscillator level'', 
$n=0\ldots \infty $. The generic state is given by
the action of this operators on the  Fock vacuum state 
$|0\rangle \tilde {|0 \rangle }$: 
\beqa \label{state} 
|a^ {\mu_1 \ldots \mu_n \mu_{n+1}\ldots \mu_{n+k} },~ p\rangle 
~=~\alpha^{\mu_1\dagger }_{m_1}
\ldots  \alpha^{\mu_n\dagger }_{m_n} 
\tilde {\alpha}^{\mu_{n+1}\dagger }_{\tilde {m}_1}
\ldots  \tilde {\alpha}^{\mu_{n+k}\dagger }_{\tilde {m}_k}
\left |0, p\right > 
\tilde {\left|0, p \right >}~, 
\eeqa 
with the constraint that the total level $N$  
of left and right oscillators be equal: 
\beqa
N~=~\sum _i~ m_i~  =~ \sum _i~ \tilde {m}_i~.
\eeqa
In Eq. (\ref {state})  $p$ is the momentum of the  state and must obey
the mass shell condition $p^2=M^2$, where $M$ is determined by
the string scale $M_{st}$ according to the Regge behavior   
\beq \label{level} 
M^2 ~=~4~(N-1)~M^2_{st}~.
\eeq
Moreover, the state (\ref{state}) must obey the transversality condition
\beq\label{transv}
p_{\mu}~|a^{\mu\ldots\mu_n\mu_{n+1}\ldots\mu_{n+k}}, p \rangle ~=~ 0~.
\eeq
Taking all possible Lorentz-irreducible combinations of indices 
the expression  (\ref{state}) 
gives rise to the states with different spins which can 
range  from $0$ (trace on all indexes) to $n+k$ 
(totally symmetric,  transverse,  traceless combination).
For example, the  states at the level $N$ are of the form 
$|a\rangle_{left} \times
|a\rangle_{right}$ with, say    $|a\rangle_{left}$,  defined as 
follows: 
\beqa \label{states} 
&\alpha_N^{\mu\dagger} \left|0 \right> \qquad &  \nonumber \\
&\alpha_{N-k}^{\mu\dagger}\alpha_{k}^{\rho\dagger}\left|0 \right> 
\qquad (k=1\ldots [N/2]) \qquad &  \nonumber \\
&\ldots \ldots &\nonumber \\
&\alpha_1^{\mu_1\dagger}\ldots \alpha_1^{\mu_N\dagger}\left|0 \right>~, 
 \qquad &
\eeqa  
and respectively for $|a\rangle_{right}$ 
with the substitution $\alpha \to \tilde{\alpha}$. 
Thus,  at each mass level  there are states of any 
spin $n$ with $ n \leq  2N $. These are given by all 
possible combination of creation operators  
whose individual level numbers add up to $N$. 
One is naturally led to the 
question: what is the total number of states with a given mass M ? 
If we forget about the Lorentz structure for a moment, this is equivalent 
of counting the total number $p(N)$  of \emph{partitions} of the 
integer $N$, i.e., the number of sets of the form $\{n_1,\ldots , 
n_k\}$\footnote{$k$
 is called the \emph{length} of the partition.}  such that 
$\sum n_i = N$. This is a well known  
problem in the number theory and the 
solution, for large $N$, scales as follows: 
\beq \label{asympt1}
p(N)~ \sim ~\exp \left ( {\sqrt{b\,N}} \right )  \qquad N\to \infty~,
\eeq 
where $b$ is a constant of order 1. Taking into 
account that each oscillator can come in 
$d=(D-2)$ varieties  ($D$ being the dimensionality of space)
the constant $b$ is 
replaced by $b\, d$. Thus, the density of 
states of a given mass $M$ grows \emph{exponentially} 
for $M>M_{st}$:
\beq \label{density}  
\rho(M) ~\sim~ \exp \left ( {\sqrt{b\,d}\ {M\over M_{st}}} \right )~.
\eeq
One possible objection against  very low quantum gravity scale is
that in string theory there is a very large number of states with masses
growing as $\sqrt{N} M_{st}$. Moreover, the number
of states at each mass level grows with 
$N$ as $(\exp{\sqrt{N}})$.
Therefore, if $\M$ is the scale where classical gravity breaks down
we should expect  to deal with the exponentially large number of 
accessible states in contradiction with experiment. 
In particular, a system with the density of states such as (\ref{density})
exhibits very peculiar  thermodynamic properties above $T=M_{st}$.
The partition function for  this system is roughly
\beq
{\mathcal Z}~ =~ \sum_E ~\rho (E) \exp \left (-{E\over T} \right )~
~ \sim ~\sum_E \exp \left ( {E\over M_{st}}~ -~{E\over T} \right )~.
\eeq
The latter diverges badly when $T\geq T_H\sim M_{st}$ 
(``Hagedorn transition'', see, e.g., Ref. \cite{bowick} 
and citations therein).

However, we will see below that what really  matters in our 
model is the number of states of a given mass produced by a \emph{given
number of creation operators} acting on the vacuum. For instance, it is 
clear from (\ref{states}) that at any level $N$ there is only \emph{one}
(modulo Lorentz permutations)
state created by two $\alpha^{\dagger}$'s (one left and one right);
In fact, it turns out \cite{andrews} that the number $p_n(N)$ of 
partitions of an integer $N$ \emph{of fixed length} $n$ 
scales for large $N$ but fixed $n$ ($n\ll N$) as follows: 
\beq  \label{fixedl}
p_n(N)~ \sim ~\frac{(N-n)^{n-1}}{n!~(n-1)!}~.
\eeq
Thus, for fixed $n$ there is only a \emph{polynomial} dependence on
$N$.  Therefore, the number of states $p_{n+k}^{(d)}(N)$ created by 
$n$ left oscillators and $k$ right oscillators 
which have a total of $n+k$ Lorentz indexes grows with $N$ at most as 
\beq \label{fixed2}
 p_{n+k}^{(d)}(N)~ \sim ~N^{n+k-2}d^{n+k}~.
\eeq
This fact will be crucial for phenomenological estimates.

\subsection{Modeling Couplings to 4D Matter} 

We now consider a five dimensional \emph{field theoretical} model
with the spectrum of closed bosonic string, namely a tower  
of  massive tensor fields $A^{C_1\ldots C_j}$, with  masses $M=
\sqrt{4(N-1)} \Ms$, and  $j\leq 2N$  for each $N=1,2,\ldots\infty$.
In particular, we associate to each string state given in  (\ref{state}) 
a tensor field with the same number of Lorentz indexes defined by 
$j=n+k$ and the corresponding Regge masses.  
The symmetric traceless part of this tensor  contains a spin-$j$ field,
the maximal spin state in the multiplet.
In addition, $A$ gives rise to lower spin states 
corresponding to its traces and/or antisymmetric 
components.
   
From the four-dimensional  point of view  each high-dimensional 
$j$-th rank tensor decomposes into various 4D fields with spins 
up to $j$. The couplings of these higher-spin fields to 4D matter
on a  brane depends on a concrete string theory realization of the model.
Since we do not really have a precise stringy model we 
take these couplings to have the following minimal form in terms of 
$A$ but a generic form in terms of the worldvolume fields:
\beq\label{interaction}
{\mathcal L}_{int}~ = ~\frac{A^{C_1\ldots C_j}(x,~y=0)} {M_*^{j - 2}}  
~\Sigma^*{\hat O}_ {C_1\ldots C_j} \Sigma 
~\equiv ~\frac{A^{\mu_1\ldots \mu_j}(x, y=0)} {M_*^{j - 2}}~  
J_{\mu_1\ldots \mu_j}(x)~,
\label{cop}
\eeq
where $\Sigma$ collectively denotes
the SM fields which are confined to the brane and thus 
do not depend on $y$, 
${\hat O}_ {C_1\ldots C_j}$ is some tensor  operator 
of dimension $j$ which contains derivatives and could also
contain  the mass $M$ of the field $\Sigma$. 
In Eq. (\ref {cop})  and below we will not be distinguishing between
symmetric and antisymmetric parts. The consideration will 
apply universally to all fields with multiple indices and scalars. 

To make  parallels with the case of a massless graviton 
we choose to work with the bulk field $A$ which is dimensionless.
In this case, the bulk kinetic term for $A$ is multiplied
by $\M^3$. Moreover, we assumed that the localized field $\Sigma$
is a scalar which has canonical 4D dimensionality, 
$[\Sigma]=[{\rm mass}]$. In the case of a 
spin-$1/2$ field, the operator ${\hat O}$ 
will also contain gamma matrices 
and will have dimensionality $j-1$. 

Therefore, we write the action for the field $A$ in the 
following form: 
\beq \label{interaction2} 
S_A ~= ~\M^3~\int {\mathrm d}^4x~dy ~\left [ (\de_{(5)} A)^2 ~-~ M_B^2 A^2 
\right]~ +~ \frac{1}{M_*^{j - 2}}\int {\mathrm d}^4 x 
~A(x, y=0)\cdot J(x)~.
\label{acA}
\eeq    
Here, $M_B$ denotes the bulk mass for the field $A$.
Below we are going to study how these high-spin fields 
affect the phenomenology  on the brane.

\subsection{Induced Kinetic Terms} \label{induced} 

Just as it happens for a graviton,  
the interaction of the tensor field $A$ with
the localized matter fields will modify its kinetic term on the brane.
In particular, the  vacuum polarization diagram 
with the internal SM lines localized on the brane (see Fig. 1) 
will in general give rise to the 
induced brane kinetic term for $A$.
Although the mechanism is very generic and could 
originate from  perturbative as well as nonperturbative 
worldvolume effects, for simplicity we will discuss below 
a one-loop effect. The expression for the corresponding diagram is:
\beq
\Gamma^{(2)}_{\mu_1\ldots\mu_n\nu_1\ldots\nu_n}(p,~y) 
~ = ~\frac{\delta(y)}{\Ms^{2(n-2)}}~\int
 {\mathrm d}^4 k \, \frac{ O_ {\mu_1\ldots \mu_n}(p,k,M)~ 
O_ {\nu_1\ldots \nu_n}(p,k,M)}{(k^2+M^2)~[(p+k)^2+M^2]}~, 
\eeq
where $M$ stands for the mass of the particle in the loop 
and the numerator in the integrand is a tensor 
of rank $2n$. This tensor is constructed out of 
the loop- and external momenta  
and the tensor $\eta_{\mu\nu}$. 
The result of the integration 
has the following generic  form (ignoring the tensor structure) 
\beq \label{general}
\Gamma^{(2)}(p,y) ~\sim~ \frac{\delta(y)}{M_*^{2(n-2)}} 
~\left[ c_1~ \msm^{2n}~  + ~c_2 ~p^2 \msm^{2n-2}~ +~\ldots ~+~ c_n~ p^{2n}
\right]~ \ln \msm,
\eeq
where $c_n$'s are some coefficients and $\msm$ denotes, 
as before, the ultraviolet cut-off of the 
world-volume  theory. Here for simplicity we 
took a particle in the loop the mass of which 
is much smaller  than $\msm$. In general this does not need to  
be the case and additional mass corrections should be included; 
however, for heaviest states, i.e.,  $M\sim \msm$,
the form of Eq. (\ref {general}) will remain the same
\footnote{Hereafter, for simplicity we will not discriminate
between $\msm$ and the induced scale $M_{\rm ind}$, i.e.,
we put $\msm \sim M_{\rm ind}=\mpl$.}. 

As a result of this diagram, 
the loop-corrected effective action for the field $A$ will 
contain additional kinetic and higher derivative terms 
which are localized on the brane. These terms arise 
respectively from the momentum dependent parts in (\ref {general}),
while the first term in  equation (\ref{general}) represents a brane 
mass term for $A$; we will discuss it momentarily. 

The second term in (\ref{general}) is equivalent to a four 
dimensional kinetic term in the action. 
The largest contribution to it 
comes from either the cutoff and/or the heaviest 
particle in the world-volume theory 
the mass of which is of order $\msm$. 

Therefore, as in the previous sections,
this leads to the coefficient in front 
of the induced 4D kinetic term which is of the order of $M_{\rm P}$.
Indeed, the induced 4D kinetic term for the higher spin field
can take the form\footnote{Here we discuss the higher spin states
with $n\ge 2$ which give rise to the dangerous exponential 
multiplicity of states. The case with $n=0,1$ will be 
discussed briefly below.}:
\beq \label{indkin}
S_{n} ~=~ \frac{M_{\rm P}^{2(n-1)}}{M_*^{2(n-2)}}~\int {\mathrm d}^4 x  
~ [\de_{(4)} A(x,~y=0)]^2~.
\label{AAA} 
\eeq 
This expression sets the  ``crossover scale'' $r_c^{(n)}$ 
for the field $A$ to be
\beq \label{rcn}
r_c^{(n)}~ = ~\frac{M_{\rm P}^{2n-2}}{\Ms^{2n-1}}~ = ~\frac{1}{\Ms}~
\left(\frac{M_{\rm P}}{\Ms}\right)^{2n-2}~. 
\eeq
Moreover, the coupling of this field to the localized matter
(analog of the Newton coupling) is:
\beq
G_{(n)}~=~{1 \over \M^{2n-1}~r_c^{(n)}}~=~{1\over \mpl^{2(n-1)}}~.
\label{cA}
\eeq
Since $\M r_c^{(n)}\gg \gg 1$, this coupling is tremendously suppressed
compared to what it would have been if we were to neglect the induced
kinetic term on the brane. Thus, we see the same phenomenon:
the SM fields shield themselves from the strong bulk dynamics.
  
Let us note that the parameter $r^{(n)}_c$ 
depends on the rank of the tensor field.  In particular, it
is determined by the dimensionality  
of the operator ${\hat O}_{\mu_1\ldots\mu_n}$ 
to which this tensor field is coupled in (\ref{interaction}).  
The  latter is related to  the number of derivatives by which the 
field $A$ couples to 4D matter. 
Above we have assumed 
that a given number $n$ in (\ref{rcn}) corresponds to 
$n$ oscillators acting on the Fock 
vacuum in the oscillator  picture.  This fact will be important 
in counting these fields with the right multiplicity.

We turn now to the first term in Eq. (\ref{general}). This 
is just a four-dimensional induced mass term for the field $A$. 
Depending on the scenario at hand, this 
term can take two significantly different values.
We will study below both of these possibilities.

For a generic  interaction the first term in Eq. 
(\ref{general}) will take the form:
\beq 
S_{m}~ = ~{\msm^{2n}\over \M^{2(n-2)}} 
~\int {\mathrm d}^4 x [A(x,~y=0)]^2~.
\eeq
After rescaling the field appropriately to bring it
to a canonical dimension we obtain that the 
4D mass of this field is $\msm$.  Therefore, it is not likely that
such a heavy state could  play any role in the low-energy 4D 
dynamics on the brane. Indeed, 
the total action for the field $A$ now consists of three parts,
$S_A + S_n + S_m$.
The latter has the form of the action which we 
discussed in section \ref{theframework} for a 
scalar field with the  bulk mass $M_B$  and the brane
mass $\mu \sim \msm $. 
According to the results obtained in  section \ref{theframework}, this 
state will effectively  appear to four-dimensional observers 
as a 4D  field of mass $\msm$; thus, it will  have no effect
on the 4D physics at accessible energies. 
Therefore, in this case there is no additional constraint on 
the value of $M_*$.

However, one could expected that for some particular choices of the 
interaction terms (\ref{interaction2}) the induced mass 
on the brane could  be  much smaller, e.g., of order 
$M_B$. To come to this point let us recall that 
for a massless  spin-two field which couples gravitationally 
to 4D matter the  4D reparametrization 
invariance  prevents the generation of any type of 
mass term. 
Moreover, in the case of a \emph{massive} spin-two state 
the bulk reparametrization invariance is explicitly broken 
by the mass term.
Thus, one could  expect that the brane mass term will be induced
by radiative corrections.  The latter, however, has to 
vanish in the limit of zero  tree-level bulk mass. Therefore,
the induced mass could in principle be determined by  
the bulk mass. 

It is reasonable to expect that this will happen 
also in the higher spin cases for some specific choice of the interaction 
current in (\ref{interaction2}).
Indeed, in the massless limit the higher form symmetric and antisymmetric
fields have the corresponding well-known gauge invariant actions.
This invariance is explicitly broken in the bulk by the mass term.
In particular, if the matter current on the brane 
$J$ is conserved up to the terms which are proportional to the 
tree level bulk mass of the field $A$, i.e.,
$\de \cdot J~ \sim {\cal O} \left ( M_B^2 \right )$,   
then the induced mass term could be of the order of the bulk mass,
$\mu \sim M_B$. In this case these light fields will have 
an interesting impact on the 4D phenomenology on the brane\footnote{This
consideration and the phenomenological discussions below  
also apply to the case of a pure four-dimensional
theory where the Regge modes  would have masses of order $\M$.}.
The content of the next section is devoted to the 
analysis of the phenomenological, astrophysical and cosmological data
which might be affected by the presence of these states.

Before we turn to these studies let us summarize briefly the 
main properties of these light modes. The total action for the field 
$A$ takes the form:
\begin{eqnarray}
S~= ~\M^3~\int {\mathrm d}^4x~dy ~\left [ (\de_{(5)} A)^2 ~-~ M_B^2 A^2 
\right]~+~\frac{1}{M_*^{n - 2}}\int {\mathrm d}^4 x 
~A(x, y=0)\cdot J(x)~ \nonumber \\
+~ \frac{M_{\rm P}^{2(n-1)}}{M_*^{2(n-2)}}~\int {\mathrm d}^4 x  
~ \left [ (\de_{(4)} A(x,~y=0))^2 ~ -~ \mu^2~ A(x,~y=0)^2 \right ]~.
\label{totA}
\end{eqnarray}
Here, as we discussed above, $\mu \sim M_B~=
\sqrt{4(N-1)}M_*$, $N = 1,2, \ldots $.
This action is of the form  discussed in section \ref{theframework}
and analyzed in detail in Appendix.
According to these results we can  
distinguish two cases: if $\mu  >  M_B$ and $\mu< M_B$.
In the first case, for each bulk field 
$A$ we find a continuum of modes with masses   
starting at $M_B$. Only a small portion of the continuum 
with the mass around $\mu$ 
and width  around $1/r_c^{(n)}$ 
is unsuppressed on the brane: 
the brane induced kinetic term converts their strong bulk couplings 
into a significantly weaker coupling $1/\sqrt{M_*^3 r_c^{(n)}}$, 
similar to what happens for the graviton \cite{dgp,dgkn}. 

To give an example, consider a state from  
the continuum with the mass $m$.
It can be produced in a process 
involving brane-fields. The amplitude, $F$, for this process 
is  proportional to the bulk-brane coupling  which 
in its turn is specified by the square of the wave-function at 
the position of the brane. Thus, we write
\beq
|F|^2~ \sim ~\frac{|\phi^{(n)}_{m,\mu}(y=0)|^2}{M_*^{2n-1}}~ 
|F^{(n)}_m|^2~,   
\eeq
where $F_m^{(n)}$ is a kinematical factor. 
In order to obtain the total cross section 
this must be integrated over $m$  from $M_B$ to $\infty$. 
However,  the function $|\phi_m(y=0)|^2$  is sharply peaked around
the value of  $\mu$ (see Appendix) which is nothing but the 
brane-induced mass. The width of this peak is of the  order of 
$1/r_c^{(n)}$. The result of the integration is as follows:
\beq
\int_{M_B}^{\infty} ~{\mathrm d} m \, 
\frac{|\phi^{(n)}_{m,\mu}(y=0)|^2}
{M_*^{2n-1}}~ |F_m|^2 ~  \sim ~\frac{1}{M_*^{2n-1}~ r_c^{(n)}}
|F^{(n)}_{m=\mu}|^2~.
\eeq
Therefore, the integration procedure 
has effectively converted the \emph{large} coupling 
constant $1/M_*^{2n-1}$ into a significantly 
suppressed constant  $  1/( M_*^{2n-1} r_c^{(n)})$, as advertised 
before. 

Furthermore, we consider the case $\mu < M_B$. 
All the states in the continuum are significantly 
suppressed on the brane. However, as discussed before, there is in
addition a localized 4D state of mass $\sim \mu$.
The coupling of this state is also suppressed by the 
parameter  $1/r_c^{(n)}$ (see Appendix).

Hence, in both cases considered above  the 
situation is identical for all the practical purposes: 
the  relevant contribution to any  4D process
comes only from the states with ``effective'' 4D mass around 
$\mu \sim M_B$. 
These states are  coupled to the 4D matter by the 
weak coupling  $1/(M_*^{2n-1} r_c^{(n)})$.    
In what follows, we will discuss for simplicity 
only the case $\mu > M_B$,  keeping in mind 
that the physics  is similar  even if $\mu<M_B$.

So far we were dealing with the massive Regge modes
for which the exponential multiplicity is present.
However, in addition we expect to have a few 5D massless modes
in the bulk. For instance, in the bosonic sector of a close string  
a graviton will be accompanied by a dilaton and a two-form antisymmetric 
field. Although these massless modes are not important for the problem
of the exponentially growing number of states, nevertheless,
they could mediate gravity competing forces in 4D worldvolume. 
Therefore, we should deal with them. The dilaton can acquire a
potential on a brane; in this case it cannot mediate  
gravity competing 4D force \cite {dil}. The higher dimensional
two-form field will give rise to a pseudoscalar and a vector
particle on the brane. The pseudoscalar can be dealt in analogy 
with the dilaton, the brane-induced potential will suppress its 
interactions. However, the massless vector field should be dealt separately.
One possibility is  to give to it a small mass by 
the Higgs mechanism. 

After these discussions we turn to the constraints.

\section{Phenomenological Constraints}

In this section we study 
the  production  of the  five-dimensional higher 
spin fields introduced in the previous section, in processes 
taking place on the brane. The resulting estimates for 
rates and cross sections  will be subsequently used to put bounds
on $M_*$ through the analysis of astrophysical, cosmological 
and  collider observations. 
 
\subsubsection{Annihilation}

A five dimensional high spin field  coupled to brane fields   can be 
produced in the processes when the localized
charged matter annihilate on the brane.

Consider the production of a ``stringy'' $n$-th rank tensor 
mode $A_m^{(n)}$ 
of ``4-dimensional mass'' $m$ (i.e., belonging to the continuum of states)
in the annihilation  process involving a massless brane fermion-antifermion
or brane scalar-antiscaler,  and a
brane-photon, $\Phi  {\bar \Phi} \to \gamma A_m^{(n)}$ . 
The amplitude has the form   
\begin{equation}
F^{n}_m~ =~\frac{e}{\Ms^{3/2+n-2}} ~\phi_m (0) ~\epsilon^{*\mu} 
\epsilon_A^{* \mu_1\ldots\mu_n}~ 
\frac{O_{\mu\mu_1\ldots\mu_n}(p,p',q)}{q^2}~,\\
\end{equation}
where $p, p'$ are the incoming  
momenta of the $\Phi$ particles, 
$q$ is the momentum transfer,  $\epsilon^\mu$ , 
$\epsilon_A^{\mu_1\ldots\mu_n}$ are the polarizations of the photon 
and of the $A$ field, respectively, and 
$O_{\mu\mu_1\ldots\mu_n}(p,p',q)$ is a 
tensor of dimension  $n+1$  whose form depends on the specific 
choice  of the interaction term (\ref{interaction}).  
Summing over initial and final polarizations we get 
an expression for the amplitude squared which can be 
used to calculate the density of the differential cross section:
\beq \label{cross}
\frac{ d^2\,\sigma_n}{d\, t~ d\,m } ~\sim~
\frac{e^2 ~|\phi_m(0)|^2}{\Ms^{2n-1}}~  s^{n-3}
~f_n(t/s, m^2/s)~,
\eeq
where $f_n(x,y)$ is a dimensionless function whose precise form depends
on the result of the sum over polarizations and on the form of the 
interaction term, and $s$ and $t$ are  the Mandelstam variables.
From the kinematics in the center of mass (CM)  frame we get
$s=4E^2$ , $t=  1/2\,(s-m^2)(\cos\Theta_{CM} -1) \in [m^2-s, 0] $,
therefore
\beqa 
{d\sigma_n \over d\,m}~ \sim ~ e^2\frac{|\phi_m(0)|^2} { \Ms^{2n-1}}~
s^{n-3}~
\int_0^{s-m^2} \,  d (-t) \,f_n(t/s, m^2/s)~  = \nonumber \\
=~e^2~|\phi_m(0)|^2 \frac{s^{n-2}} {\Ms^{2n-1}}~ 
\int_0^{1-z}~f_n(x,z)\,  dx~ ,
\eeqa
where $z\equiv m^2/s$.
Now we must integrate over $m$. As we discussed  above, due to the form 
of $|\phi_m(0)|^2$, this effectively 
amounts of  replacing $m$ by $\mu$
and $\Ms^{2n-1}$ by $r_c^{(n)}\Ms^{2n-1}$.  
The result is suppressed by powers of $\mpl$:
\beq \label{comptoncross} 
\sigma_n ~\sim~ e^2~
\frac{s^{n-2}} {r_c^{(n)} \Ms^{2n-1}} ~\int_0^{1-z}
f_n(x,z)\,  dx~=~e^2~ \frac{s^{n-2}} { \mpl^{2n-2}} 
~\int_0^{1-z} f_n(x,z)\,  dx~.
\eeq

\subsubsection{Photoproduction}

Another type of process  which contributes to the production of five 
dimensional higher spin modes is 
the photoproduction reaction $\Phi \gamma \to \Phi A_m^{n}$
with the bulk field $A$  in the final state. 

The photoproduction  rate   into   the bulk modes can be calculated
in the same way as it was done for the annihilation process. 
The cross section is the same
as in (\ref{cross}), with   $s$ and $ (-t)$ 
interchanged in the amplitude. Thus  the  density of the differential 
cross section will take the form
\beq
\frac{ d^2\,\sigma_n}{d\, t~d\,m }~ \sim~ 
e^2~\frac{|\phi_m(0)|^2}{\Ms^{2n-1}}~(-t)^{n-3}~
~g_n(t/s, m^2/s)~,
\eeq
where $g(x,z)$ is another dimensionless function.  
Integrating this expression over the continuum with the approximation
$|\phi_{m}(0)|^2\sim\delta(m-\mu)/r_{c}^{(n)}$,  
we get 
\beq \label{ann}
\frac{ d\,\sigma_n}{d\, t}~ \sim~ \frac{e^2}{r_c^{(n)}\Ms^{2n-1}}
~s^{n-3}~ \left ( {(-t)\over s} \right )^{n-3} ~g_n(t/s, \mu^2/s)~=~
\frac{e^2}{\mpl^{2n-2}}
~s^{n-3}~{\tilde g}_n(t/s, \mu^2/s)~,
\eeq
where we have introduced ${\tilde g}_n\equiv ( {(-t)/ s})^{n-3} ~g_n$.
This expression for the differential cross section 
is again suppressed by powers of $\mpl$.

\subsection{Star Cooling} 

The possibility of producing an exponentially large number of
Regge states at very low energy 
could in principle affect the cooling rate of stars and supernovae.
Requiring that this effect be smaller than the observed energy produced
by these objects (in particular SN1987) was indeed the strongest
constraint on the model introduced in \cite{add}, as was also
shown in \cite{perels}. In this case   
the states in question were KK modes of the graviton, 
with mass spacing of the
order of 1 mm$^{-1}$. Let us consider the total production  
of bulk   modes by a stellar object of temperature
$T$. 
For example, we can estimate  the emission rate due to 
the 'photoproduction' 
or 'bremsstrahlung' processes. As far as the order of magnitude is 
concerned, the other processes considered in \cite{add}, \cite{perels}
will give roughly the same contribution. 

The rate is given by  the cross-section (\ref{comptoncross}) 
multiplied by  the number of particles per unit
volume ($\sim T^3$) , times the relative velocity of the initial 
particles, everything averaged over the thermal bath of the star. 
This gives   
\beq \label{gamma} 
\Gamma_n \sim  E \left(\frac{E}{M_{\rm P}}\right)^{2n-2}, 
\eeq
in this expression  $E\equiv \left<E\right> \sim T$.
This must be multiplied by  
the number of  states with given $n$ that contribute 
to the rate. As discussed in the previous section, 
this is equal to the number of states which, 
in string language, are created by $n$ oscillators. 
On the other hand, the number of such 
states at each level  $N$ goes at most 
as  in  Eq. (\ref{fixed2}), with $n$ 
in place of $n+k$.
 
Thus, the total rate of the 
production of all the states specified by a given 
$n$ but belonging to an arbitrary mass level is then bounded 
as follows (neglecting the factor of $d^n$)
\beq
\Gamma  ~\leq~   \sum_{N=1}^{N_{max}}   E \left(\frac{E}{M_{\rm P}}\right)
^{2n-2} N^{n-2} 
~\sim~  E ~\left(\frac{E}{M_{\rm P}}\right)^{2n-2} N_{max}^{n-1}~,
\eeq
where $N_{max}$ is determined by the mass of the heaviest state that
 can be produced at energy $E$, i.e. 
$\mu_{max} \simeq \sqrt{N_{max}} M_* = E$. 
Thus we get the estimate
\beq
\Gamma \sim   E \left(\frac{E}{M_{\rm P}}\right)^{2n-2} 
\left(\frac{E}{\Ms}
\right)
^{2n-2} \sim E \left(\frac{E^2}{M_{\rm P}\Ms}\right)^{2n-2}.
\eeq
Even for  $\Ms\sim$ mm$^{-1}$, the 'natural' expansion parameter,
which is $E^2/(M_{\rm P}\Ms)$, 
becomes of order 1 at energies above 1 TeV. Thus, 
the rates for the different $n$'s start being of the same order at 
a scale much higher than that at which all these states are accessible
by  kinematics. 
Therefore, in a star where $T\ll$ 1 TeV, all contributions other than 
$\Gamma_2$  are negligible:    
\beq \label{gamma2} 
\Gamma_2 ~\sim ~\sum_{N} \frac{E^3}{M_{\rm P}^2} ~ \sim ~\frac{E^3}{M_{\rm 
P}^2}~
N_{max}~ \sim~ E~\frac{E^4}{\Ms^2 M_{\rm P}^2}~.
\eeq

This rate is strikingly similar 
to that of the model of Ref. \cite{add} 
for the case of \emph{two}  large extra-dimensions, 
where $\Ms$ plays the role of $1/R$. So the lower bound on $\Ms$ from 
this kind of processes is  precisely of the order of an inverse 
millimeter or so.     
   
\subsection{Collider Signatures}

The similar considerations apply to collider
experiments. Although a huge number of 
states can be produced starting at a very
low energy ($\sim \Ms$), the cross sections 
for production of states labeled by different 
$n$   do not  become significant until 
energies 
of order $\sqrt{M_{\rm P}M_*}(\sim 1$ TeV if $M_*\sim 10^{-3}$ eV).
 Thus, the dominant contribution at smaller  
center of mass  energies 
 comes from $n=2$, and according to Eq. (\ref{ann}) it scales
as follows:  
\beq
\frac{ d\,\sigma_2}{d\, t}~ \sim ~ \frac{1}{s M_{\rm P}^{2}}~.
\eeq
As before,  we have to sum over the $n=2$  states 
of each mass level  up to $N_{max} \sim s/M_*^2$. The result is 
\beq\label{totalcross}
\frac{ d\,\sigma_2}{d\, t} ~\sim ~ \frac{1}{M_{\rm P}^2 M_*^2}~.
\eeq
For the comparison, 
the contribution of the states with  a generic $n$ is given 
(at most) by (see (\ref{ann}))
\beq\label{totalcross2} 
\frac{ d\,\sigma_n}{d\, t}~ \sim ~\sum_{N=1}^{N_{max}} 
\frac{s^{n-3}}{ M_{\rm P}^{2n-2}}~ N^{n-2}~ \sim ~\frac{1}{s^2}
~\left(\frac{s}{ M_{\rm P} M_*}\right)^{2n-2}~ ,
\eeq
in which we find again that $s/( M_{\rm P} M_*)$ is 
a natural expansion parameter. Since
the denominator in this ratio is at least (1 TeV)$^2$, 
therefore, at the energies accessible  
in present day colliders only the contribution
from $n=2$ will be important.  
 
For  $n=2$   the cross section (\ref{ann})  is of the 
same form as the one calculated in Ref. \cite{ratt}, where  
the production of  massive spin 2 KK modes in models 
with large extra dimensions of size $R$ was considered.
The  expression found in \cite{ratt} is:   
\beq\label{ratta}
\frac{ d\,\sigma}{d\, t} ~\sim~    \frac{1}{s M_{\rm P}^{2}}~ 
F_1(t/s,\mu^2/s)~, 
\eeq
where the precise form of $F_1$ is given in \cite{ratt}. 
In this case, one must sum over the tower of KK modes corresponding to 
$n$ 
compact extra dimensions, with  masses  given by 
\beq\label{KK}
m^2~= ~\frac{1}{R^2}~ (k_1^2 +\ldots + k_n^2)~,
\eeq
 for  integers $k_i$, $i=1 \ldots n$,    
so the number of available states up to 
energy $s$ is roughly $(\sqrt{s}R)^n$.  In particular, 
for the case of two extra 
dimensions, we get from (\ref{ratta}) 
\beq
\sum_{KK \, modes} \frac{ d\,\sigma}{d\, t} ~\sim ~
\frac{R^2}{M_{\rm P}^2}~.   
\eeq
Remarkably, this is the same result we found in 
(\ref{totalcross}) 
provided  that we exchange $M_* \leftrightarrow 1/R$ 
(although the two frameworks are totally 
different). 
In particular, the bounds obtained in \cite{ratt} on $R$ 
by comparing their results with present 
 collider data, can be directly translated 
into  bounds on our $M_*$, which therefore  is again  only  
constrained to be larger than $\sim $ $1/$mm.
 
Despite of this similarity, 
the predictions of our model and the one considered in 
\cite{ratt} begin to differ drastically when the energy is high enough: 
indeed, when the bound is saturated, the framework  with two large extra 
dimensions predicts the existence of 
the  density of states which grows roughly linearly with the energy
\beq
\rho(m) ~\sim~ m ~R^2~,
\eeq
while in the present context all states with $n\geq 2$ 
will become equally 
important (the 
dimensionless ratio in (\ref{totalcross2}) becomes of order 1), 
and the total  density of states is \emph{exponentially} increasing
\beq
\rho (m) \sim  e^{m/M_*}~d(m). 
\eeq
(Here $d(m)$ is some function whch contains 
powers of $m$ and depends on dimensionality of space.)
Therefore, for instance, the spectrum of missing  
energy signatures will be very  different in these two cases.

\subsection{Cosmology}

In this section we will consider cosmological constraints coming from
the overproduction of bulk states.
In order to be as model independent as possible,
we shall discuss the following initial
conditions for the hot big bang:

 (1) The bulk is virtually empty;

 (2) The brane states are in thermal equilibrium at some
temperature $T_{brane}$.

Our goal is to find out what is the normalcy temperature $T_*$ defined in
\cite{add} as the temperature below which Universe expands as normal
4D FRW Universe. Then, by requiring that $T_*$ be at least higher than
the nucleosynthesis temperature,
we can derive bounds on $M_*$. The reason why we
expect that this requirement may restrict $M_*$ is  the fact that
the brane can cool by ``evaporation'' into the bulk string states. If this
rate is
higher than the cooling rate due to the expansion, the FRW scenario will
be affected. We do not want this to happen below
the nucleosynthesis temperatures.
This may impose some constraints on $M_*$.

\vspace{0.1in}

{\it Cooling by Evaporation into Bulk String States}

\vspace{0.1in}

 In order to estimate the rate of bulk state production at temperature
$T$ we shall use the star cooling rate (\ref{gamma2})
where we shall substitute $E$ by $T$.
This tells us that at each mass level the
dominant contribution to the cooling rate comes from the production
of 2-index fields, provided the natural ``expansion parameter''
$T/\sqrt{M_{\rm P}M_*}$ is smaller than one. 
For $M_* \sim 10^{-3}$eV this requires
$T$ to be  below TeV. Then the cooling rate is given by
\begin{equation}
 \Gamma_2(T)~ \sim~ {T^5 \over M_{\rm P}^2M_*^2}~.
\end{equation}
 The resulting
change
of the matter energy density on the brane
due to evaporation is
\beq
 {d\rho\over dt}|_{evaporation} \sim - T^4
\Gamma_2(T)
\sim - {T^9 \over M_{\rm P}^2M_*^2}~.
\label{coolrateu}
\eeq
This has to be compared with the cooling rate caused by the
cosmological expansion
\beq
 {d\rho\over dt}|_{expansion} \sim
-3H\rho\sim -3{T^2\over M_{\rm P}}\rho~,
\label{coolexpansion}
\eeq
where $H$ is the  Hubble parameter.
In the radiation dominated epoch of standard FRW cosmology ($H \sim
T^2/M_{\rm P},\rho\sim T^4$),
the ratio of the two rates is
\beq
 {{d\rho\over dt}|_{evaporation} \over  {d\rho\over dt}|_{expansion}}~
\sim~ {T^3\over M_{\rm P}M_*^2}~.
\eeq
Requiring this ratio to be $\ll1$ we find that, for $M_* \sim 10^{-3}$
eV, $T_* \sim$ 20 MeV  or so. Thus these 
considerations put approximately the same
bound on $M_*$ as colliders experiments 
and astrophysics: for lower  $M_*$  the normalcy temperature is not  high
enough for standard nucleosynthesis  to
proceed unaffected. On the other hand, the 
 cosmological evolution above 
this  scale is dramatically modified 
and requires independent study.

\section{Black Holes}
\setcounter{equation}{0}

The sources localized on the brane
at distances $r <r_c$ interact via the weak four-dimensional gravity.
On the other hand, the sources in the bulk interact via 
strong five-dimensional gravity. This fact will  have  
interesting implications for  the black hole physics. 

Let us consider an elementary particle of mass $M$ such that 
$M_{\rm P}\gg M \gg M_*$. In a  crude approximation
we can think of it as a gravitating source
of  uniform density localized within its Compton wave-length
$\sim 1/M$. From the point of view
of the brane observer this particle is not a black hole. 
However, the very 
same particle in the bulk would appear as a black hole
since its 5D Schwarzschild radius is bigger thank its Compton wavelength
(see below). Thus,  if such a  particle
is gradually removed from the brane 
it turns into a bulk black hole. 

We shall  investigate how
the transition between the brane gravity to the bulk gravity 
takes place. For this purpose, 
we will study first the bulk gravitational potential between
two object of masses $m_1$ and $m_2$ (see Fig.3).

\vspace{5mm}
\centerline{\epsfig{file=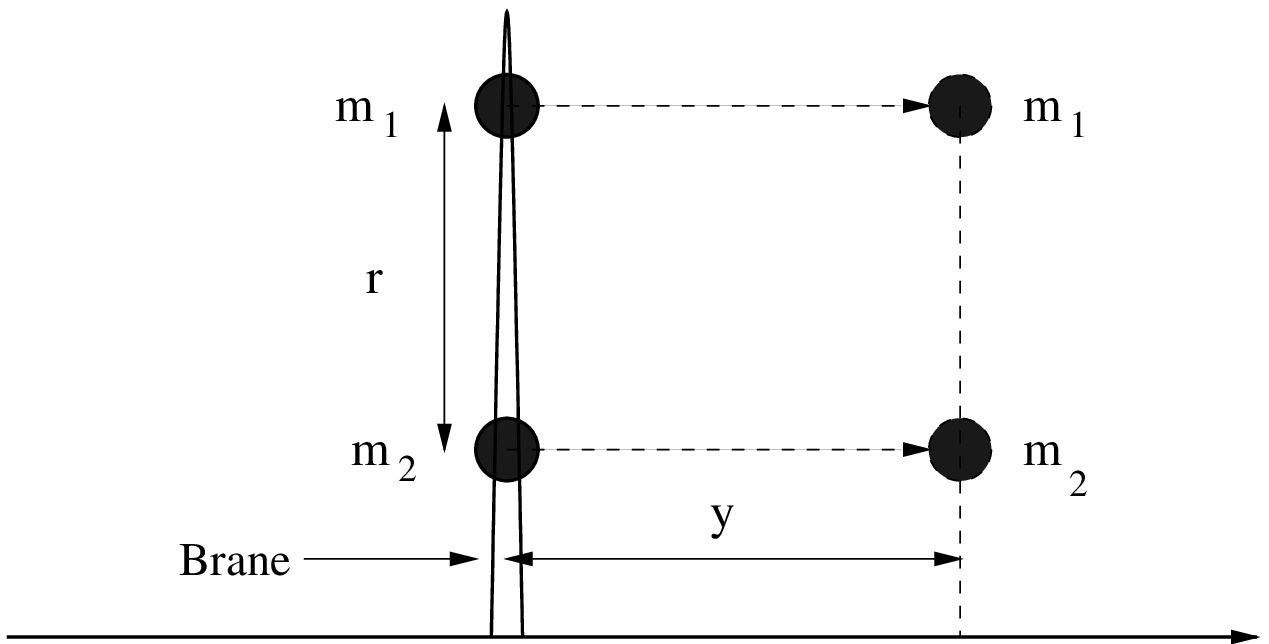,width=10cm}}
{\footnotesize\textbf{Figure 3:} Removing the test masses
from the brane into the bulk, 'switches on' the   five-dimensional 
potential.
For $r<r_{c}$ the five-dimensional potential is stronger then the
four dimensional one.}
\vspace{5mm}

\subsection{Interactions on the Brane and in the Bulk}

The full 5D static potential can be obtained by summing 
over the continuum of KK modes. Each of these can be viewed as
a four-dimensional massive particle which  generates a Yukawa-type force
between the
sources. The superposition of these 
contributions  gives the following potential:
\beq\label{trlevpot}
V(r,y)=-\frac{m_1~m_2}{16\pi^2~M_{*}^3}~\int_{0}^{\infty}
|\phi_{m}(y)|^2~\frac{\exp(-mr)}{r}~dm~.
\eeq
Inserting  the explicit form of the wavefunctions $\phi_m$
for the KK modes
(\ref{fi1},\ref{ap2d}) we find the expression for the potential to be
\begin{eqnarray}\label{trlvpot1}
V(r,y)&=&-\frac{m_1m_2}{M_{*}^3}\int_{0}^{\infty} dm
\Big(\frac{[2~\cos(my)~+~m~r_{c}~\sin(my)]^2}{4~+~m^2~r_{c}^2}
\Big)~\frac{\exp(-mr)}{r}~.
\end{eqnarray}
We can approximately evaluate this integral by dividing the 
range of integration into the two regions, $m<1/r_c$ and
$m>1/r_c$. Since $r_c$ is the largest scale, we can take 
both $r/r_c$ and $y/r_c$ to be  much smaller than unity. Let us  look at 
the value of the integrand in the first region. The  exponential  can
be replaced  with unity.  As a result, 
the contribution to the integral equals to
$1/(rr_c)$. For  $m>1/r_c$ 
one can again evaluate the integral approximately 
which in this case is equal to $2y^2/(r^4+4r^2y^2)$.
Thus, the approximate expression for  the potential is 
\beq\label{totalpot}
V(r,y)\approx-\frac{m_1~m_2}{16\pi^2~M_{*}^3}~\left(\frac{1}{rr_c}~+~
\frac{2y^2}{r^4~+~4r^2y^2}\right)~.
\eeq
We will see below that this  agrees with an exact expression
for the potential which we will obtain from the propagator. 

It is not difficult to interpret this expression. For $y=0$ we see
the ordinary Newton  potential governed by 
$G_N=1/16\pi M_{\rm P}^2 \sim 1/(r_c M_{*}^3)$
(note that we look at distances $r<r_c$). 
After the sources are moved off the
brane the strong potential which  is not suppressed by $r_c$ is 
switched on. For $y<r$ the correction to the 
potential is $(y/r)^2(1/r^2)$, while for $y>r$
a ``full-strength'' five-dimensional potential, $1/\M^3 r^2$, 
is recovered.

Let us study the more general case when the two sources
are placed at  different positions $y$ and $y_0$ in the extra coordinate.
Instead of using the KK picture we will directly solve for the five
dimensional (Euclidean) propagator 
\beq\label{gfp}
\left(\Box_4(1~+~r_{c}~\delta(y))~-~\partial_{y}^2~\right)~G(x-x_0,y,y_0)
~=~\delta^{4}(x-x_0)~\delta(y-y_0)~.
\eeq
The Green's function depends separately on $y$ and $y_0$, since 
five dimensional translational invariance is broken by the presence
of the brane. By Fourier-transforming this
expression with respect to $x-x_0$ and going to the Euclidean momentum
space $E\longrightarrow ip_4$, 
(i.e. $p^2\longrightarrow -p^2$ and $p\equiv \sqrt{p^2}$)  
we find the following equation:
\beq\label{eucprop}
\left((p^2~-~\partial_{y}^2)~+~p^2r_{c}~\delta(y)\right)~
{\tilde G}(p,y,y_0)~=~\delta(y-y_0)~.
\eeq
This equation can be solved with the ansatz
\beq\label{gfans}
{\tilde G}(p,y,y_0)~=~A(p,y_0)~e^{-p|y|}~+~B(p,y_0)~e^{-p|y-y_0|}~,
\eeq
where $A(p,y_0)$ and $B(p,y_0)$ are the functions to be determined. 
Inserting Eq.  (\ref{gfans}) into  Eq.  (\ref{eucprop}) we find 
\beq\label{AB}
A(p,y_0)~=~\frac{-e^{-p|y_0|}}{p+1/r_c},\quad\quad B(p)~=~\frac{1}{p}~,
\eeq
where we have used the identities $\partial_{y}|y|\equiv \epsilon(y),\quad 
\partial_y
\epsilon(y)=2~\delta(y),\quad \epsilon(y)^2=1$.
The momentum space Euclidean Green  function can be written as
follows:
\beq\label{gfwa}
{\tilde G}(p,y,y_0)~=~\frac{1}{p}~e^{-p|y-y_0|}~-~\frac{1}{p}~e^{-p(|y|+|y_0|)}~
\frac{1}{1+1/r_cp}~.
\eeq
Since the quantity $pr_c$ is  large  (we are considering interaction
at distances $r \ll r_c$) we can expand the denominator of
the second term to get the expression
\beq\label{expgf}
{\tilde G}(p,y,y_0)~\simeq ~\frac{1}{p}~e^{-p|y-y_0|}~-~\frac{1}{p}~
e^{-p(|y|~+~|y_0|)}~+~\frac{1}{p^2r_c}~e^{-p(|y|~+~|y_0|)}~.
\eeq
The Fourier transform of this Green  function is the potential
between two static sources of mass $m_1$ and $m_2$ at positions
$y$ and $y_0$ in the fifth dimension and separated by the
distance $r$ along  the brane worldvolume. 
By straightforward integration
we find the expression for the potential
\beq\label{coorsppo}
V(r,y,y_0)\simeq -\frac{m_1m_2}{16\pi^2M_{*}^3}\left(
\frac{1}{r^2+|y-y_0|^2}-\frac{1}{r^2+(|y|+|y_0|)^2} 
+\frac{1}{rr_c}\arctan\frac{r}{(|y|+|y_0|)}\right).
\eeq
The potential (\ref{coorsppo}) reveals some interesting properties.
For instance, if the masses are placed on different sides of the brane,
or if one of the masses is located  on the brane, the first two terms 
cancel exactly. Since the first two terms correspond to the strong
five-dimensional potential (coupled with  $1/M_{*}^3$), objects on opposite
sides  of the brane (Fig.4) interact only via the third term in 
(\ref{coorsppo}), which corresponds to the weak 4D gravity. 

\vspace{5mm}
\centerline{\epsfig{file=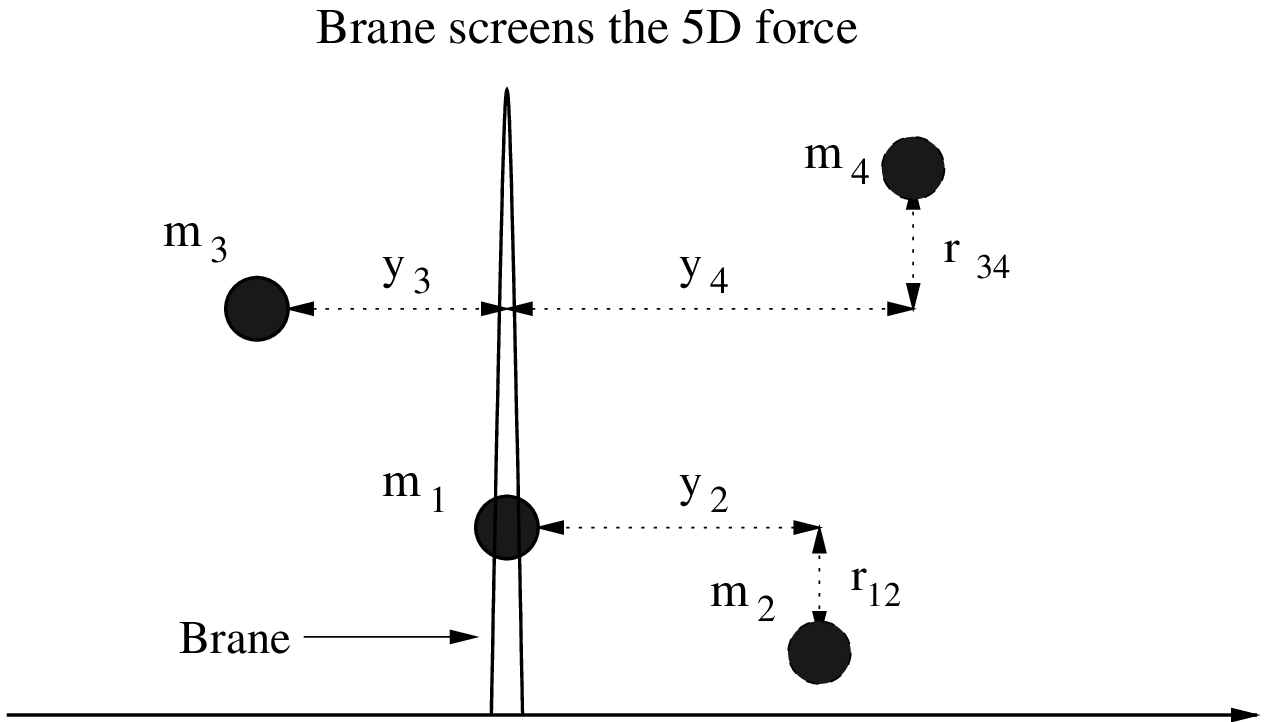,width=10cm}}
{\footnotesize\textbf{Figure 4:} The brane suppresses the
exchange of higher KK modes. Object placed on different sides
of the brane (for  example $m_3-m_4$) interact only with the third term
in (\ref{coorsppo}). The same is true if  one of the
objects is on the brane while the other one is in the bulk (for example
$m_1-m_2$). }
\vspace{5mm}

This term results from the exchange of KK modes with masses
$\lesssim 1/r_{c}$. Modes with $m\lesssim 1/r_{c}$ can be thought to 
form a resonance state which  mimics the exchange of a
single zero-mode graviton
coupled via the four-dimensional Newton  constant.
We conclude that the brane in some sense ``screens'' 
the five-dimensional force, 
i.e., the force due to the exchange of KK excitation of mass larger  than
$1/r_c$ is suppressed. 
This fact is clear  from the mode expansion picture.
Heavy modes are suppressed on the brane; the  Schr\"odinger equation 
(\ref{feq2}) that determines 
the wave-function profiles in the extra dimension is
just the equation
for a particle in a   one-dimensional delta function type potential
with strength proportional to the mass of that  particle. 
Therefore the contribution of heavy modes in the exchange is 
suppressed and the force is mostly due to the exchange of 
the modes with the small mass.

\subsection{Emission of Black Holes in the Bulk}
\setcounter{equation}{0}

As mentioned above, a particle of mass $M>M_{*}$ becomes a black hole 
in the bulk.
This  can be understood from the expression for the
Schwarzschild radius of the black hole in five dimensions 
\beq\label{schwr}
r_{S5}~\sim~\frac{1}{M_*}~\sqrt{\frac{M}{M_*}}.
\eeq
For $M>M_*$ the Schwarzschild radius becomes larger  than the
characteristic size of the particle $1/M$ (the Compton wavelength)
and the particle becomes a black hole. For instance, for $M_* \sim 
10^{-3}$ eV and  black holes of the 
masses $1$ eV, $1$ TeV and $10^{19}$ GeV, 
the Schwarzschild radii would equal to  
$3$ cm, $10^{4}$ m and $10^{12}$ m, respectively. 

The lifetime of a five dimensional black hole, which decays via the 
Hawking radiation,  is given by the relation
\beq\label{lifetbh}
\tau_{5}~\sim~\frac{1}{M_*}~ \left( \frac{M}{M_*}\right)^2~,
\eeq
and it is substantially larger than the lifetime of a 
four-dimensional black hole with the same mass
($\tau_{4}\sim M^3/M_{\rm P}^4$).
For  $M_* \sim 10^{-3}$ eV , the  lifetimes of a black hole  
the with masses of 
1 eV, 1 TeV and $\mpl$  would be $10^{-4}$ s,
$10^{19}$ s and $10^{50}$ s, respectively. 

One may wonder if there is a possibility that a heavy bulk-particle is 
produced on the brane (e.g. in an accelerator) and is emitted in the bulk 
and after becoming a
long-lived  black hole, is attracted back to the brane where it decays.
Such an event could produce an interesting signature of a displaced 
vertex\footnote{Here we are discussing the particle 
which is not necessarily localized on the brane.}. 
Unfortunately the probability of such an event is very low as we shall 
briefly discuss.

To determine the relative rate for the events in which a particle
emitted into the bulk returns back to the brane within the size of the 
detector we have to evaluate
the relevant fraction of the phase space.
Let us assume that a particle of bulk mass $M$ is produced on the brane in a 
process of energy $E$. If we denote the magnitude of the momentum along 
the brane by $p$ and the momentum in the
transverse direction by $p_y$ then 
\beq\label{1steq}
p^2+p_{y}^{2}\le E^2-M^2~.
\eeq
The constraint that the particle comes back to the brane can be
expressed in terms of its escape velocity from the brane 
(which in our case can be estimated as $v_{esc}\simeq 10^4~$ m/s): 
\beq\label{2ndeq}
|p_y|\le Mv_{esc}~.
\eeq
We also require the particle to be within the detector when
it hits the brane. This constrains the maximum value of the 
momentum $p$ along the brane. During the motion in the $y$
direction the particle experiences an approximately constant force
due to Earth's gravity, with the acceleration $g=10~m/s^2$
\beq\label{fy}
F(r=R_{E},y)=-\frac{\partial V}{\partial y}=
-\frac{Mm_{E}}{M_{*}^{3}R_{E}r_{c}}\partial_{y}\arctan(r/y)=
-\frac{Mm_{E}}{M_{*}^{3}r_{c}}\frac{1}{y^2+R_{E}^2}\simeq -Mg.
\eeq
The time needed
for the particle to return 
back is $t=2p_y/Mg$. If we take the radius of the detector
to be $l$, this translates into the condition
\beq\label{3rdeq}
|p|\le\frac{M^2lg}{2p_y}~.
\eeq
The fraction of  phase space for which the black holes come back to the  
brane within the distance $l$ from the place of production can easily be 
visualized from the plot in the $p-p_y$ plane (Fig.5).

\vspace{5mm}
\centerline{\epsfig{file=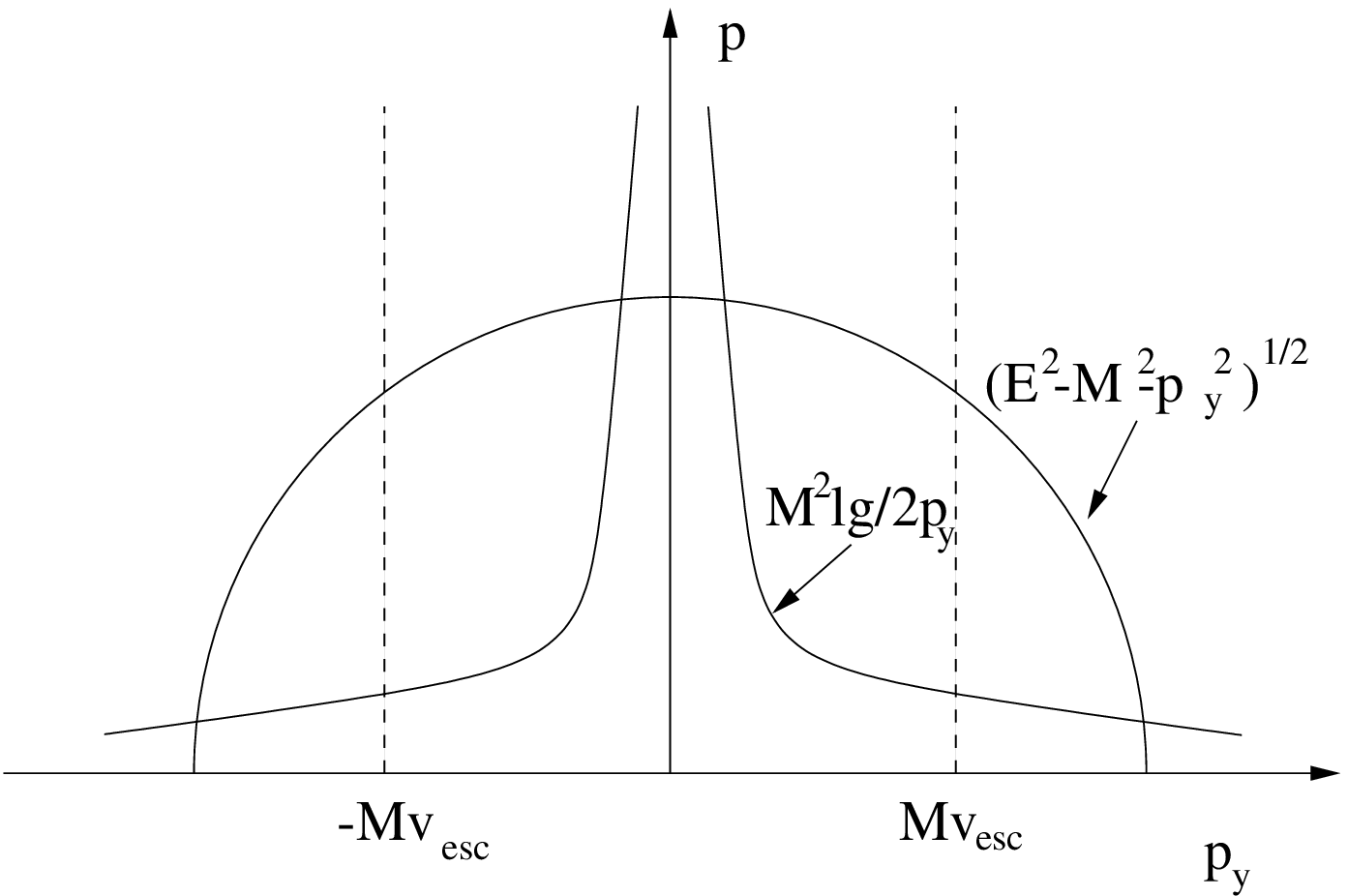,width=10cm}}
{\footnotesize\textbf{Figure 5:} The phase space for the production
of  black holes. $E$ is the energy available in the experiment.  
Conservation
of energy requires that $p^2+p_{y}^2\le E^2-M^2$ (semi-circle). 
For a  black hole to hit the brane, $p_y$ must satisfy $p_y\le Mv_{esc}$
(vertical line). 
In order to hit the brane within a detector of radius $l$, the 
momentum $p$ must
be less then $M^2lg/2p_y$. The  ratio of the area 
satisfying these constraints
to the area of the semi-circle is the phase space suppression factor.}
\vspace{5mm}

Let us first estimate the branching ratio for the black hole produced at 
TeV energies to come back to the brane. 
We have to estimate what will
be the fraction of the 'Regge' states that will not have significant
phase space suppression for falling back to the brane due to attraction
by the Earth's gravity. Those are particles that satisfy 
$M^2v_{esc}^{2}\ge (E^2-M^2)$, i.e.,
\beq\label{br1}
0<\frac{1}{n}\left(\frac{1\mbox{TeV}}{10^{-3}\mbox{ev}}\right)^2-1
\lesssim 10^{-9}~.
\eeq
Solving (\ref{br1}) for $n$, we find that for 
$n\in[10^{30}-10^{21},10^{30}]$
(when $M_{*}=10^{-3}$eV)
there is no phase space suppression\footnote{Note that 
the particles with such a large $n$ can be black holes
from the standpoint of the worldvolume theory as well.}. 
The  ratio  of 
the total rate $\Gamma_{tot}$ of  black hole production,
to the  rate 
$\Gamma_{back}$ for production of those  
that will return to the brane 
is
\beq\label{br2}
\frac{\Gamma_{back}}{\Gamma_{tot}}=10^{-9}.
\eeq
Some of the black holes that  come back to the brane would be attracted
toward the center of the earth and would not be detected. 
Let us now 
estimate the number of  black holes that will hit the brane within 
a detector of radius $l$, which  can be 
taken to be of order one meter or so.
Solving for the intersections of (\ref{1steq}) and (\ref{3rdeq})
one finds that for 
\beq\label{br3}
0<\left(\frac{E^2}{M^2}-1\right)<\frac{lg}{c^2},
\eeq
the two curves do not intersect, i.e., the  semi circle is fully contained
within the curve (\ref{3rdeq}) ($c$ is the speed of light). 
When the condition (\ref{br3})
is satisfied, there is no phase space suppression for the black holes 
to come back to the 
brane within the detector size. Let us translate the bound
(\ref{br3}) into the bound on the 
number of Regge states at $E=1$TeV that have no
phase space suppression. One finds that the ratio of the total number
of kinematicaly accessible modes to the 
 number of modes that don't have phase 
space suppression is just $lg/c^2$. Therefore,
the ratio of the rate  for a 
black hole entering back into the detector,  
$\Gamma_{det}$, to  the total rate  
of  black hole production $\Gamma_{tot}$ is
\beq\label{br4}
\frac{\Gamma_{det}}{\Gamma_{tot}}=\frac{lg/c^2}{1+lg/c^2}\approx lg/c^2=
10^{-16}.
\eeq
This  ratio  is practically unobservable.

\section{Baryon Number Violation by Virtual Black Holes}

\def\pf{M_{\rm Pf}}

The potential danger for any theory with a low quantum gravity scale $M_*$
are the high-dimensional operators that may violate exact or approximate global
symmetries of SM (such as flavor or the baryon ($B$) and 
lepton ($L$)
numbers). In this section we shall argue that
in our framework
the strength of such dangerous operators is suppressed 
by the scale $M_P$, and
not by $M_*$, and, therefore, they are harmless.

 In order to see this let us first discuss the possible origin of such
operators.
It is believed usually that non-perturbative quantum gravity effects
such as the virtual black holes (VBH)
violate the global symmetries of the theory.
Such a non-conservation should be seen in an effective low energy
theory as a variety of $B$- or $L$-violating effective operators, e.g.,
such as
\begin{equation}
  qqql~,
\label{dang}
\end{equation}
where $q$ and $l$ are quark and lepton fields respectively.
The question is the strength of these operators.
This issue is impossible to address without the knowledge
of the microscopic quantum gravity theory.
Nevertheless, in certain cases one can estimate the maximal strength
in a quasi-classical approximation.
The  main reason for an expectation that  VBH violate global charges
is  the no-hair theorem\cite{nohair},
which implies that BH are characterized by ``charges'' that are
coupled to the massless fields. Conservation of such charges
cannot be violated by BH,
since an outside observer can measure the conserved flux.
Such a measurement is impossible for a global charge, which
renders it uncontrollable.

In the literature one may find a number of
estimates for VBH-mediated $p$-decay first 
discussed by Zel'dovich
\cite{zeldovic}. The main idea is that an elementary
particle, carrying a global charge in question, may quantum mechanically
collapse into
a VBH, or be captured by one. VBH can later
decay into a final state of an arbitrary global charge leading to its
non-conservation. 
To estimate the rate of such a process
Zel'dovich used a ``geometric'' cross section of
the gravitational capture of a particle inside a VBH, which is
simply given by its Schwarzschild radius squared. Thus
\begin{equation}
\sigma \sim \left ({M_{BH}^2 \over M_P^4}\right )~.
\end{equation}
The resulting estimate for the proton lifetime was
\begin{equation}
\tau_p \sim {1\over m_{\rm proton}}\left ({M_P \over m_{\rm
proton}}\right )^4~.
\end{equation}
Somewhat different estimate can be obtained \cite{dolgov} if the amplitude
of the proton collapse into a VBH is evaluated from the BH - proton
wave-function overlap integral
\begin{equation}
\int d^3r \Psi_{\rm proton}\Psi_{BH}~.
\end{equation}
The main point in all these studies, most important for the
present discussion, is that either proton or
some of its constituent quarks must be trapped inside a VBH . Once
captured by a VBH
the memory about the baryon charge is erased and VBH can decay into
an arbitrary  kinematicaly  allowed set of particles with the same
color and the electric charge, but different baryon number. 
The resulting $p$-decay rate is suppressed by the powers of  
$\left ({m_{\rm proton}/ M_P}\right )$. 
Therefore, the dangerous operators (\ref{dang})
appear suppressed by powers of $M_P$
\begin{equation}
 {E^{n} \over M_P^{n + 2}}~qqql~,
\end{equation}
where $E$ is the energy in the process. Thus, in conventional
$4$D theories, were $M_* = M_P$, the corresponding rate is very much
suppressed
even for $n=1$. Naively, one may think that in theories with low quantum
gravity scale the relevant scale to be used in the
above equation instead of $M_P$ is $M_*$. This would be an obvious
phenomenological
disaster for our framework. Fortunately, this naive expectation
is wrong as we shall now
explain. Let us again consider a $B$-violating process induced by
VBH in which a proton, or some of its constituent quarks, 
collapses into a VBH. 
The relevant suppression in such a process is $M_P$, not $M_*$, due to the
fact that the strong bulk gravity is shielded from proton. Recall, that
in the limit $M_P\rightarrow \infty$ gravity switches off regardless of
the value of $M_*$.

An alternative simple way to see this is to remember that proton
is localized on the
brane, where gravity is week and the scale of a microscopic VBH is $M_P$
just as in ordinary $4D$ gravity. As a result, the collapse of a proton
into a VBH will go as in the ordinary case. The same would be true regarding
any other process that break the global symmetries of the SM. 
Thus, we conclude that
in our framework the VBH-mediated processes are harmless.

So far our analysis was done in the minimal case, in which there are
no new exotic states that can carry baryon number into the bulk. If such states
are introduced, some experimentally interesting possibilities may open up;
such are neutron-anti-neutron oscillation (without observable
proton decay). The existing bounds on such
processes are much milder than that for the proton decay, 
and they can be a subject
of an independent experimental search \cite{kamyshkov}.

For instance, a neutron may mix to a heavy bulk fermion $X$, to which we can
prescribe a baryon number $B=1$ (but zero lepton number).
However, since in the bulk the $B$-number
is not conserved due to very low mass of VBH, this mixing can lead to a
process
\begin{equation}
n \rightarrow X \rightarrow {\rm VBH} \rightarrow \bar{n}~.
\end{equation}
Note that if the corresponding mixing operator on the brane
is induced by gravity, it
will be suppressed by powers of $M_P$
\begin{equation}
 {udd{\bar X} \over M_P^{5/2}}~,
\end{equation}
and will be practically unobservable. Of experimental interest 
is the case when it is induced by non-gravitational effects, for instance, by
integrating out some {\it perturbative} heavy brane states with masses
$M \ll  M_P$, in which case the strength may be controlled by their mass
\begin{equation}
 {udd{\bar X} \over M^{5/2}}~,
\end{equation}
and can be experimentally observable. Note that the source of the baryon number
non-conservation, is again gravitational, since the above operator {\it per se}
does not violate baryon number. $B$-violation can only occur, if the virtual
$X$-fermion collapses into a bulk BH. Note that the analogous high-dimensional
operator for proton
will also require more SM particles in the final state as well as
violation of the lepton number and thus
will be suppressed by additional powers of $M_P$.

\section{Conclusions}

In this paper we proposed a framework in which the ``rigid'' SM
is coupled to gravity which becomes ``soft''
above the scale $M_* \ll  \msm $. 
It was assumed that the quantum theory of gravity
above $M_*$ has some generic properties of the closed string
theory. We showed that the bound $M_* > 10^{-3}$ eV is compatible
with all the present day observations, despite the exponentially
increasing density of string states. The key phenomenon 
is ``shielding'' by which the rigid SM makes gravity weak 
without affecting its softness. 
This is due to the renormalization of the kinetic term of
a graviton and other string states by SM loops.
As a result, the 4D gravitational coupling is set by 
{\it non-gravitational} physics, 
while the scale of the softness is still determined by
the gravitation. 

We discussed an explicit model
in which the SM lives on a 3-brane embedded in 
infinite-volume flat 5D space.
The spectrum of 5D bulk gravity above the scale $M_*$ is that of a 
closed string theory. 
Despite this fact, a brane observer sees at the distances $M_*^{-1}\ll r \ll 
M_P^2/M_*^3$ the 4D gravity with the Newton constant set by 
the SM physics. 

In high energy processes on the brane the production of the string states
becomes significant only at energies above $E \sim \sqrt {M_P~M_*}$.
As a result, collider experiments, astrophysics and early cosmology
{\it independently} put the same lower bound,  $M_* \sim 10^{-3}$ eV. 
The same bound is obtained from sub-millimeter gravitational 
measurements \cite{measurements}.
For this value, the model has experimental signatures both for colliders 
as well as for sub-millimeter gravity measurements.

We have discussed some unusual properties of the black holes in the
present framework. Despite the low quantum gravity scale, the 
virtual black hole mediated baryon number
violating operators are suppressed by powers of 
$M_P$ and are harmless.

If supplemented by low-energy supersymmetry, our framework maintains
the successful prediction of the gauge coupling unification 
\cite{UNIFICATION}, despite the very low value of the 
quantum gravity scale.

\vspace{0.2cm}
\begin{center}

{\bf Acknowledgments}
\vspace{0.1cm} \\

\end{center}

We would like to thank I. Antoniadis, N. Arkani-Hamed,
M. Luty, M. Perelstein, R. Sundrum and A. Vainshtein for 
useful discussions.
The work of GD was supported in part by David and Lucille Packard
Foundation Fellowship for Science and Engineering, by Alfred P. Sloan
foundation fellowship and by NSF grant PHY-0070787. 
The work of GG  is supported by  DOE Grant DE-FG02-94ER408.
The work of MK was
supported in part by David and Lucille Packard Foundation.
FN is supported by the NYU McCraken Fellowship.

\vspace{0.2in}

\section{Appendix}
\setcounter{equation}{0}

Let us consider  a five-dimensional scalar field $\Phi$,
with an induced  kinetic term on a four-dimensional brane, placed
in the origin of the fifth dimension. 
We denote the coordinates with  $x^{A}\equiv(x^{\mu},y)$. The
range of the fifth  coordinate is $y\in[-\infty,\infty]$. Equally well 
this
can be thought of as having the 
compact fifth dimension of very large radius $R$.
We will consider two separate cases,   in which the field
$\Phi$ is massless or  massive, respectively. 

\subsection{Massless field}

The model Lagrangian is 
\beq \label{L5}
{\mathcal L}~=~ \partial^A \Phi \partial_A \Phi~ +~ r_{c}~\delta(y)~ 
\partial^\mu
 \Phi\partial_\mu \Phi~ .
\eeq
The kinetic term on the brane is induced with  strength $r_{c}$.
We look for solutions 
of the form   $\Phi= \phi_m(y)\sigma_m(x^\mu)$, where    
$\sigma_m(x^{\mu})$
satisfy  the four dimensional Klein-Gordon equation
$(\partial^\mu \partial_\mu \, +\, m^2)\sigma_m =0$. Then the 
profiles $\phi_m(y)$ are determined by  the  equation
\beq\label{feq2}
\left( \partial_y^2~+~m^2 ~+~ r_c~\delta(y)~m^2 \right)~ 
\phi_{n}(y)~ =~0~.
\eeq
Outside the origin the  solutions are plane waves of frequency $m$. The 
wavefunctions have definite parity and we will concentrate on those
that have non-zero value at the origin. Let us divide the space in
two regions $ I\equiv[-\infty, 0]$, $II\equiv[0, \infty]$. We take 
the wavefunctions to be
\begin{eqnarray}\label{fi1}
&\left.I\right)& \qquad   \phi_I(y)= A \cos(my) - B\sin(my)~, \nonumber \\
&\left.II\right)& \qquad   \phi_{II}(y)= A\cos(my) +B\sin(my)~.
\end{eqnarray}
Integrating equation  (\ref{feq2}) from $y=-\epsilon$ to $y=+\epsilon$, 
we find
\beq\label{ap1j}
B^2~=~\frac{r_{c}^2m^2}{4}~A^2~,
\eeq
Since we are dealing with plane-wave-normalizable 
wavefunctions, we can choose 
\beq \label{norst}
A^2+B^2 =1/2\pi~,
\eeq
(this choice reproduces the correct normalization of the 
propagator if we use $|\phi_m(0)|^2$ as the spectral density in eq. 
(\ref{KL})). 
The resulting  value of the 
modulus squared of the  wavefunction on the brane is 
\beq\label{ap2d}
|\phi_{m}(y=0)|^2~
=~\frac{1}{2\pi}~\frac{4}{4+r_{c}^2m^2}~.
\eeq
The suppression of the squared modulus of the wavefunction is shown in the
Fig.6.

\vspace{5mm}
\centerline{\epsfig{file=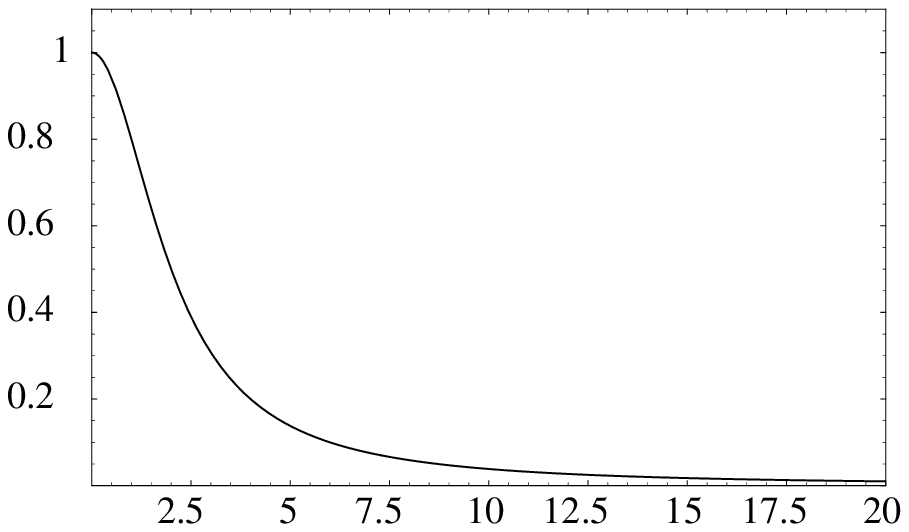,width=10cm}}
{\footnotesize\textbf{Figure 6:} Modulus squared of the wave function
at the origin (\ref{ap2d}) as a function of the mass. The 
modulus  squared is plotted on the y axis (in units of  $1/2\pi$)
, while the x axis  is the mass of the KK modes in units of $1/r_{c}$}.
\vspace{5mm}

To summarize, a massless five-dimensional field with an induced  kinetic
term on the brane gives rise to a  continuous KK spectrum, starting from
the zero mass. Higher KK states are suppressed on the brane according
to (\ref{ap2d}).

The Euclidean propagator for this model, in the case of a source located
at the origin,  is easily found as in \cite{dgp}. The defining equation 
is 
\beq 
(\Box_5 ~+~ r_c \delta(y) \Box_4)~ G(x,y)~ = ~\delta^4(x)~ \delta(y)~.
\eeq
Fourier transforming in the $x^\mu$ variables we get 
\beq \label{prop}
( p^2~-~\pr_y^2~ +~ r_c~ p^2 ~\delta(y) )~ \tilde{G} (p, y)~ =
~ \delta(y)~,
\eeq
where $p^2$ is the Euclidean four-momentum. With the ansatz
$\tilde{G} (p, y)= D(p, y)B(p)$, with  $D(p, y)$ satisfying 
$( p^2-\pr_y^2)D(p, y)=  \delta(y)$, it is straightforward to 
obtain the solution 
\beq
\tilde{G} (p, y)~= ~{1\over 2p~+~r_c p^2}~\exp\{-p|y|\}~.
\eeq

\subsection {Massive field}

Let us now consider a field  of bulk-mass $M$ and brane-mass $\mu$. 
The  Lagrangian  is
\beq \label{L5M}
{\mathcal L}~=~ \partial^A \Phi \partial_A \Phi~ + ~r_{c}~\delta(y)~ 
\partial^\mu
 \Phi\partial_\mu \Phi~ -~M^2\Phi^2~-~\mu^2~r_{c}~\delta(y)~\Phi^2~,
\eeq
where  we have suppressed an overall factor of $M_*^3$. The  
equation of motion
for the filed $\Phi$ is
\beq \label{field eq1}
\left(\partial^A \partial_A + r_c \delta(y) \partial^\mu \partial_\mu +
M^2+r_{c}\mu^2\delta(y)\right)\Phi =0.
\eeq
Decomposing the field $\Phi$ in  KK modes we end up with the
equation
\beq\label{feq3}
\left( \partial_y^2+ r_c\delta(y)(m^2-\mu^2) \right) \phi_{m}(y) =
(M^2-m^2)\phi_{m}(y).
\eeq
This is the equation for a  particle of energy $m^2-M^2$ in
a delta-function type potential. For $m>M$ there will
be a continuum of scattering states; moreover, 
in the case $\mu<M$,   for $m< M$  there
are also  two bound states in the spectrum. 
Let us first look at  the case $m<M$.
 The solutions in regions I and II   are 
\beqa\label{fi2}
&\left.I\right)& \qquad   \phi^{BS}_{I}(y)= A\exp(\sqrt{M^2-m^2}y)~, 
\nonumber \\
&\left.II\right)& \qquad   \phi^{BS}_{II}(y)= A\exp(-\sqrt{M^2-m^2}y)~.
\eeqa
Integrating equation (\ref{feq3}) from $y=-\epsilon$ to $y=+\epsilon$
gives the condition
\beq\label{condA}
2\sqrt{M^2-m^2}=r_{c}(m^2-\mu^2).
\eeq
This has no solution for $\mu>M$, while for $\mu<M$ it  is satisfied for
\beq\label{bsma}
m_{BS}^{2}=\mu^2-\frac{2}{r_c^2}\pm\sqrt{\frac{1}{r_{c}^4}
+\frac{M^2-\mu^2}{r_{c}^2}}.
\eeq
The  modulus squared of the  bound state
wavefunction at the origin  is easily evaluated from the 
normalization
condition
\beq\label{norco}
A^2\int_{-\infty}^{\infty}\exp(-\sqrt{M^2-m^2}|y|)dy=1=
A^2\frac{2}{\sqrt{M^2-m^2}}.
\eeq
We can derive an effective four dimensional action for the localized mode 
by integrating the Lagrangian (\ref{L5M})  over the fifth dimension, 
writing
$\Phi = \phi_b (y) \sigma_b(x)$, with $\sigma_b$ dimensionless and 
$\phi_b =\exp(-\sqrt{M^2-m^2})|y|$
: 
\beqa \label{Leff}
{\mathcal L}_{b} & = &  M_*^3 \Bigg\{\sigma_b \Box \sigma_b \left[\int 
{\mathrm d} y \,\phi_b^2 \bigg(1  + r_c \delta(y)\bigg)\right]  \nonumber 
\\
 & + &   \sigma_b^2\left[\int{\mathrm d} y \,\phi_b \bigg( - \pr_y  + M^2 
+
 \mu^2 r_c \delta(y) \bigg) \phi_b   \right]\Bigg\}~,
\eeqa
where we have restored the 
factor $M_*^3$. Using  equation  (\ref{feq3}) for  $\phi_b$, we get 
\beq  
{\mathcal L}_{b}=  M^3_*~\left(r_c + \frac{1}{\sqrt{M^2-m_{BS}^2}}\right)
\left[\sigma_b \Box \sigma_b + m_{BS}^2 \sigma_b^2\right]~.
\eeq
As usual,  the overall coefficient is  the inverse square of the coupling
constant  of this  four dimensional mode to  matter. Using
(\ref{bsma}) we can write it as 
\beq 
G_{b}^{-1} = M_*^3 r_c\left[ 1 -  \frac{1}{1\pm\sqrt{1+r_c^2(M^2-\mu^2)}}
\right] \simeq M_{\rm P}^2 \left (1 + O(1/M r_c)\right )~.
\eeq
Therefore, the localized modes are coupled with strength  $M_{\rm P}$.     
Now let us look at the continuum of states with $m>M$. The  wavefunction
for these  modes is 
\begin{eqnarray}\label{contwf}
&\left.I\right)& \qquad   \phi_{I}(y)= A \cos(\sqrt{m^2-M^2}y) - 
B\sin(\sqrt{m^2-M^2}y)~, \nonumber \\
&\left.II\right)& \qquad   \phi_{II}(y)= A\cos(\sqrt{m^2-M^2}y) +
B\sin(\sqrt{m^2-M^2}y)~.
\end{eqnarray}
Once again, matching the derivatives at the origin gives the condition
\beq\label{wf3}
2\sqrt{m^2-M^2}B=r_{c}A(m^2-M^2).
\eeq
From (\ref{wf3}) and the condition (\ref{norst})  
we can find the modulus squared of wavefunction at zero
\beq\label{resona1}
|\phi_{m}(y=0)|^2= \frac{1}{2\pi}\left [ {1~+~\frac{r_{c}^2 m^2}{4}
\left(\frac{(1-\mu^2/m^2)^2}{1-M^2/m^2}\right)}\right ]^{-1}~.
\eeq
The term $1/(1-M^2/m^2)$ in the denominator ensures  that the
wavefunction at the origin vanishes for $m=M$. For $\mu<M$ the suppression
is much like the suppression in (\ref{ap2d}). 
When $\mu$ becomes bigger than $M$, the  bound state disappears and
the continuum states with mass $m\approx \mu$ are enhanced on the
brane. The enhancement of continuum modes with mass close to $\mu$ is
shown in Fig.7.

\vspace{5mm}
\centerline{\epsfig{file=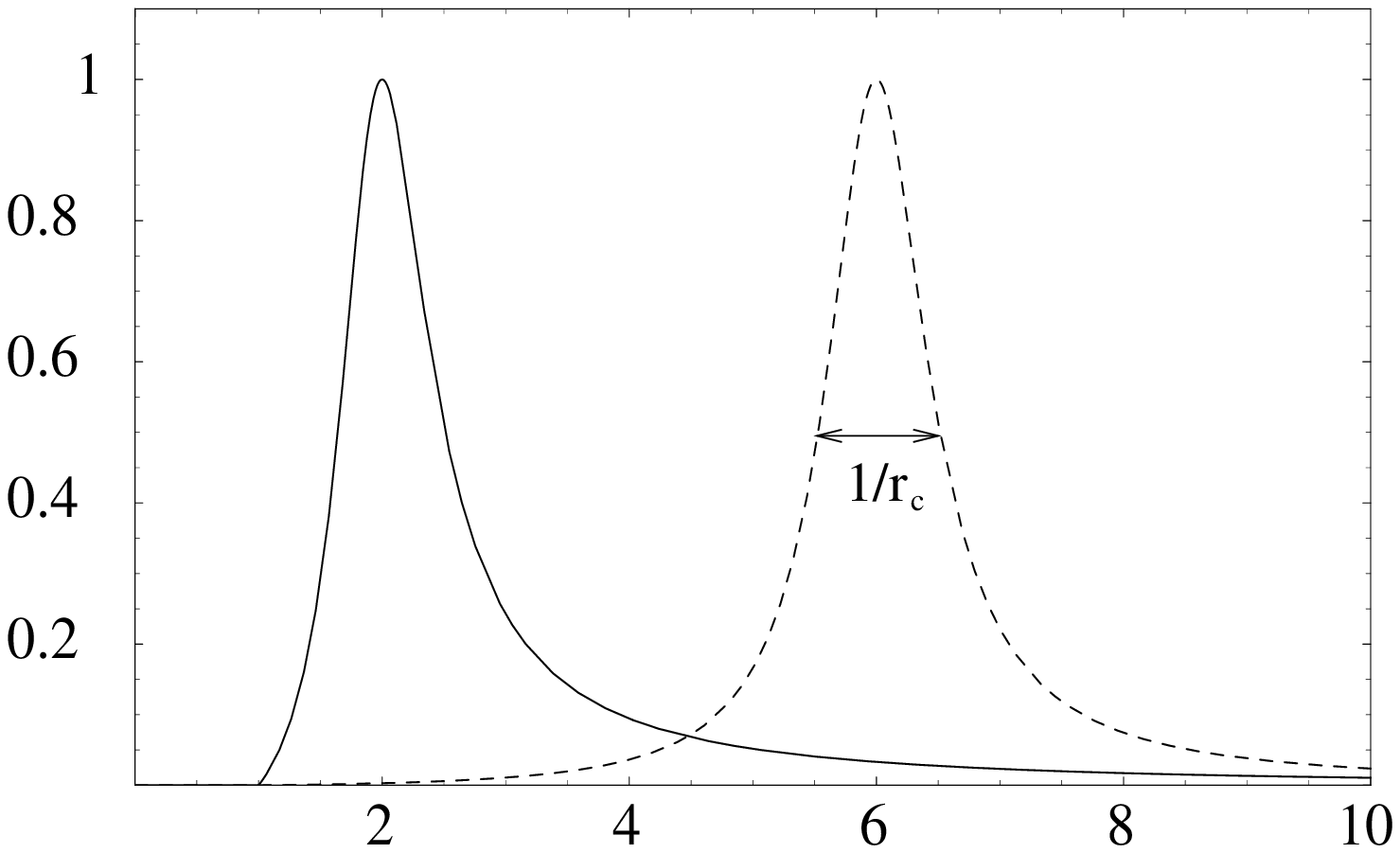,width=10cm}}
{\footnotesize\textbf{Figure 7:} 
Modulus squared of the wavefunction on the brane, for the
continuum modes and two choices $\mu=2/r_c$ (solid), $\mu=6/r_c$
(dashed), and  $M=1/r_c$. 
The modulus  squared is  plotted on the y  axis (units are $1/2\pi$).
On the x axis is the mass $m$ of the KK modes in units of $1/r_{c}$. }
\vspace{5mm}

The suppression (\ref{resona1}) can be rewritten in terms of
the dimensionless variable $x\equiv mr_{c}$ like
\beq\label{sup22}
|\phi_{m}(0)|^2~=~\frac{1}{2\pi}~ \left [ {1~+~\frac{1}{4}
\left(\frac{(x^2-\mu^2r_{c}^2)^2}{x^2-M^2r_{c}^2}
\right)}\right ]^{-1}~.
\eeq
The dimensionless function in (\ref{sup22}) (also shown on Fig.7 for
a specific
choice of values $M$ and $\mu$) has extremal value one at $x=\mu r_{c}$. 
The
width of the peak,  defined as the distance between points at which the 
function drops to one half of its maximal value,  is $\delta x\approx 
1+\frac
{M^2}{4\mu^2}$. For any function $f(m)$ which is slowly varying over a 
range 
of the order $1/r_c$, we can then make the approximation
\beq \label{approx}
\int dm |\phi_m(0)|^2\, f(m) \simeq \frac{1}{r_c} f(\mu)~.
\eeq
This approximation is  extensively used in the text.   
Each bulk mode of mass $M<\mu$ produces  a continuum of KK states
with masses  $m>M$. Due to the specific resonance form of the wavefunction
on the brane for those modes, we can approximate their effect
with a single mode of mass $\mu$. This  'mode' is coupled to
brane fields with  strength proportional to $1/r_{c}M_*^3$.  
The Euclidean  propagator can be easily found also in the massive case:
equation (\ref{prop}) is replaced by    
\beq
[ p^2-\pr_y^2 + M^2 + r_c  \delta(y)(p^2 + \mu^2  ) ]~\tilde{G} (p, y)~ =
~ \delta(y)~.
\eeq
Again, write $ \tilde{G} (p, y) = D(p,M,y)B(p,M,\mu)$ with $D(p,M,y)$
satisfying $(p^2+M^2-\pr^2_y)D(p,M,y)=\delta(y)$. It is then 
straightforward
to obtain 
\beq 
\tilde{G} (p, y)= \frac{1}{r_c}{1\over 2\sqrt{p^2+M^2}/r_c 
~+~(p^2+\mu^2)}~
\exp\{-\sqrt{p^2+M^2}|y|\}~.
\eeq
Notice that the above expression has a pole  corresponding to
the bound state found above. Also, notice that for large momenta compared
to $1/r_c$, and for $y=0$, it becomes approximately 
\beq
\tilde{G} (p, y)\simeq \frac{1}{M_*^3 r_c} \frac{1}{p^2 + \mu^2}~,
\eeq
where we have reinserted the appropriate overall factor of $1/M_*^3$ that 
should multiply the Lagrangian (\ref{L5M}). This 
describes a four dimensional state  with mass $\mu$ and coupled with  
strength $1/M_{\rm P}^2$, in agreement with the previous discussion.     

\subsection{Short-Distance Potential in String Theory}

In this Appendix we show how  string theory softens the 
behavior of the gravitational interaction
between two static sources, preventing the  potential from 
blowing up in the short-distance limit. 
 
For simplicity, consider closed bosonic string theory in the
critical dimension $D=26$. 
The interaction potential between two static point-like sources
of masses $m_1$ and $m_2$   separated 
by a distance $r$ can be found\footnote{We take 
these sources to couple to string states similar to D0-brane couplings.} 
in \cite{polchinski}:
 \beq \label{cilinder}
 V(r)~=~ m_1 m_2 ~\int_0^\infty \frac{d t}{t} t^{-1/2}
 \exp \left( \frac{-t \, r^2} {2\pi \alpha'} \right)
  \left(\eta(it)\right)^{-24},
\eeq
where $\eta(\tau )$ is the Dedekind $\eta$-function: 

\beq\label{eta}
\eta(\tau) = \left( \exp \{ 2\pi i \tau \} \right)^{1/24} 
\prod_{n=1}^\infty (1- \exp \{ 2\pi i  n \tau \}). 
\eeq
This function has the property $\eta(-1/\tau) = (-i\tau)^{1/2} 
\eta(\tau)$  and has the following expansion: 

\beqa 
 \eta(i \tau)^{-24} & = & \exp(2\pi \tau) + 24 + 
O(\exp(-2\pi \tau))  \qquad \qquad \; \; \;  \tau \to  \infty 
\label{asympt}\\
 \eta(i \tau)^{-24} &  = &  \tau^{12} \Big[\exp(2\pi/\tau) + 24 
+ O(\exp(-2\pi/\tau)) \Big]  \qquad  \tau \to 0. \label{asympt2}
\eeqa   
The behavior of  (\ref{cilinder}) for large separation compared to the 
string scale $\sqrt{\alpha'}$ is readily obtained noting 
that the integral is dominated by the small-$t$ region:  
using (\ref{asympt2})
(and dropping the first term in the expansion, which corresponds to 
tachyon exchange and is unphysical) we find that 
 at large distances the potential behaves as $m_1 m_2 / r^{23}$, 
reproducing  Newton's law in 26 space-time dimensions. 
However, we are interested in the short distance behavior (
$r^2 \ll 
\alpha'$). In this regime the whole integration domain contributes, 
and we cannot just insert one of the expansions 
(\ref{asympt}),(\ref{asympt2}) 
in place of $\eta(it)$. Nevertheless, we can reason as 
follows: suppose the potential blows up as $r \to 0$, as does 
any potential mediated by point-particle exchange; then we should
expect the integral  to diverge when we put $r=0$ in 
(\ref{cilinder}). 
This however is not the case: the integral can only diverge at the
extrema, and by  (\ref{asympt}) and  (\ref{asympt2}) 
\beqa
   & \frac{1}{t^{3/2}}
  \left(\eta(it)\right)^{-24} \sim  24\,  t^{21/2} & \qquad 
t \to 0   \nonumber \\   
& \frac{1}{t^{3/2}}
  \left(\eta(it)\right)^{-24} \sim 24\,  t^{-3/2} & \qquad 
t \to \infty , 
\eeqa
where we have again dropped the first  term in the expansion of 
$ \left(\eta(it)\right)^{-24}$. We see that the 
integration is finite at  both  extrema,  meaning that the potential 
does not diverge at $r=0$, and 
has therefore an expansion of the form 
\beq
V(r) = a_0   + a_1 r + O(r^2)~, \qquad \qquad r \to 0. 
\eeq
In other words, the stringy behavior of gravity, 
which becomes relevant  for distances below the string length, 
smooth out the divergences characteristic of interactions mediated
by point particles, and ``softens'' the short-distance behavior of 
gravity. This can be expressed by saying that the graviton propagator
has a form-factor that becomes effective 
above the fundamental Planck scale and makes   the 
ultraviolet behavior milder.

\end{document}